\documentclass[12pt,a4paper]{article}

\usepackage[colorlinks=true, linkcolor=black!50!blue, urlcolor=blue, citecolor=blue, anchorcolor=blue]{hyperref}
\usepackage[font=small,labelfont=bf,margin=0mm,labelsep=period,tableposition=top]{caption}
\usepackage[a4paper,top=1.5cm,bottom=1.7cm,left=2.5cm,right=2.5cm,bindingoffset=0mm]{geometry}

\usepackage{graphicx,ragged2e}
\usepackage{afterpage}
\usepackage{epsfig,cite,ulem}
\usepackage{amssymb}
\usepackage{amsmath}
\usepackage{bm}
\usepackage{dsfont}
\usepackage{multirow}
\usepackage{url}
\usepackage{xcolor}
\usepackage{ulem}
\usepackage{url}
\usepackage{booktabs}
\usepackage{tabularx}
\def\gsim{\mathrel{\rlap{\lower4pt\hbox{\hskip1pt$\sim$}}
    \raise1pt\hbox{$>$}}}       
\def\lsim{\mathrel{\rlap{\lower4pt\hbox{\hskip1pt$\sim$}}
    \raise1pt\hbox{$<$}}}       
\def\lapprox{\lower .7ex\hbox{$\;\stackrel{\textstyle <}{\sim}\;$}}
\def\gapprox{\lower .7ex\hbox{$\;\stackrel{\textstyle >}{\sim}\;$}}

\newcolumntype{C}[1]{>{\centering\arraybackslash}p{#1}}

\numberwithin{equation}{section}
\numberwithin{figure}{section}
\numberwithin{table}{section}

\begin{document}

\begin{flushright}
Nikhef/2020-021\\
\end{flushright}
\vspace{0.3cm}

\begin{center}
  {\Large \bf The Strangest Proton?}
\vspace{1.4cm}

Ferran Faura$^{1}$,
Shayan Iranipour$^{2}$,
Emanuele R. Nocera$^{1}$,
Juan Rojo$^{1,3}$,\\[0.2cm] and Maria Ubiali$^{2}$

\vspace{1.0cm}
 
{\it \small

$^{1}$Nikhef Theory Group, Science Park 105, 1098 XG Amsterdam, The
  Netherlands \\[0.1cm]  
$^{2}$DAMTP, University of Cambridge, Wilberforce Road, Cambridge, \\
  CB3 0WA, United Kingdom \\[0.1cm]
$^{3}$Department of Physics and Astronomy, VU,
    1081 HV Amsterdam, The Netherlands

}

\vspace{1.0cm}

{\bf \large Abstract}

\end{center}
We present an improved determination of the strange quark and anti-quark
parton distribution functions of the proton by means of a global QCD analysis
that takes into account a comprehensive set of strangeness-sensitive
measurements: charm-tagged cross sections for fixed-target neutrino-nucleus
deep-inelastic scattering, and cross sections for inclusive gauge-boson
production and $W$-boson production in association with light jets or charm
quarks at hadron colliders. Our analysis is accurate to
next-to-next-to-leading order in perturbative QCD where available, and specifically includes
charm-quark mass corrections to neutrino-nucleus structure functions. We find
that a good overall description of the input dataset can be achieved and that
a strangeness moderately suppressed in comparison to the rest of the light
sea quarks is strongly favored by the global analysis.

\clearpage
\tableofcontents

\section{Introduction}
\label{sec:introduction}

An accurate determination of the strange quark and anti-quark parton
distribution functions (PDFs) of the proton ~\cite{Gao:2017yyd,Kovarik:2019xvh,
  Ethier:2020way} is key to carrying out precision phenomenology at current and
future colliders, specifically for measuring fundamental parameters of the
Standard Model (SM) such as the mass of the $W$ boson~\cite{Aaboud:2017svj}, the
Weinberg angle~\cite{Sirunyan:2018swq}, and electroweak parameters in general~\cite{Bagnaschi:2019mzi}.
Because of the limited experimental information available, however, the strange
quark and anti-quark PDFs remain much more uncertain than the up and
down sea quark PDFs.

The strange quark and anti-quark PDFs have been determined from neutrino-nucleus
deep-inelastic scattering (DIS) for a long time, specifically from
measurements of dimuon cross sections, whereby the secondary muon originates
from the decay of a charmed meson, $\nu_{\mu}+N\to\mu+c+X$
with $c\to D\to \mu+X$~\cite{Bazarko:1994tt,KayisTopaksu:2008aa,
  Mason:2007zz,Samoylov:2013xoa}. When interpreted in terms of the ratio between
strange and non-strange sea quark PDFs,
$R_s\equiv(s+\bar{s})/(\bar{u}+\bar{d})$,
these measurements favor values around $R_s\lsim 0.5$ when PDFs are
evaluated at values of the momentum fraction $x=0.023$ and
scale $Q=1.6$~GeV. Therefore, it came as a surprise when a QCD analysis
of the $W$- and $Z$-boson rapidity distributions measured by the ATLAS
experiment in proton-proton collisions~\cite{Aad:2012sb}, later
corroborated by an analysis based on an increased integrated
luminosity~\cite{Aaboud:2016btc}, suggested instead a
ratio closer to $R_s\simeq~1$. Complementary
information on the strange quark and anti-quark PDFs is provided by $W$-boson
production in association with light jets~\cite{Sutton:2019pug} and
charm quarks~\cite{Stirling:2012vh}, the latter process being dominated by the
partonic scattering $g+s \to W+c$. Measurements of these processes were
performed by the ATLAS~\cite{Aaboud:2017soa,Aad:2014xca} and
CMS~\cite{Chatrchyan:2013uja,Sirunyan:2018hde} experiments recently.
Although ATLAS and CMS $W$$+$$c$ measurements turned out to be consistent
at  the
parton level, different interpretations in terms of $R_s$ were 
claimed~\cite{Aad:2014xca,Chatrchyan:2013uja}.

This state of affairs has motivated studies of the proton
  strangeness within the CT, MMHT, and NNPDF global fits, with overall
consistent findings.
The NNPDF3.1 analysis~\cite{Ball:2017nwa} found
that, whereas the ATLAS $W$, $Z$ dataset~\cite{Aaboud:2016btc} does indeed favour a
larger total strangeness, its $\chi^2$ remains 
non-optimal when fitted together with the neutrino dimuon data. 
The recent CT18 global
 analysis~\cite{Hou:2019efy} also
 presented fits with and without the ATLAS
 measurement of~\cite{Aaboud:2016btc}, with the resulting PDFs
 differing by more than one-sigma both
  for the gluon and for the total strangeness.
 An update of the global PDF analysis from
 the MMHT collaboration~\cite{Thorne:2019mpt},
 which for the first time accounted for the NNLO massive corrections to
 the neutrino dimuon cross-sections
  within a PDF fit, also revealed an enhanced strangeness driven by the
  ATLAS $W$, $Z$ dataset. The resulting PDFs were however consistent within uncertainties
  with the corresponding fit  once this dataset was excluded.
Additional dedicated studies of the strange quark and anti-quark
PDFs have been presented~\cite{Lai:2007dq,Kusina:2012vh,Alekhin:2014sya,
  Alekhin:2017olj,Cooper-Sarkar:2018ufj}, however these focused on a
restricted set of processes or datasets, or were based on theoretical and
methodological assumptions that can potentially bias the results.

 Given its phenomenological relevance for precision physics at the LHC,
a global reinterpretation of all of the strangeness-sensitive measurements
within an accurate theoretical and methodological framework appears to be
therefore  timely and compelling.   This paper
fulfills this purpose: we present an improved determination of the strange
quark and anti-quark PDFs, accurate to next-to-next-to-leading order (NNLO)
in perturbative QCD  where available, by expanding the NNPDF3.1 analysis~\cite{Ball:2017nwa}
in two respects. First, we take into account several new pieces of experimental
information which are relevant to constrain the strange quark and anti-quark
PDFs: charm-tagged to inclusive cross section ratios measured by the NOMAD
experiment~\cite{Samoylov:2013xoa} in fixed-target neutrino-nucleus DIS;
and an extended set of cross sections for inclusive gauge-boson production and
$W$-boson production in association with light jets or charm quarks measured by
the ATLAS~\cite{Aaboud:2016btc,Aad:2014xca,Aaboud:2017soa} and
CMS~\cite{Chatrchyan:2013uja,Sirunyan:2018hde} experiments in proton-proton
collisions. Second, we improve the theoretical description of dimuon neutrino
DIS structure functions, by implementing NNLO charm-quark mass corrections, and
of $W$$+$$c$ production data, by including a theoretical uncertainty that
accounts for the unknown NNLO QCD corrections; we also explicitly enforce the
positivity of the $F_2^c$ structure function.

The outline of this paper is as follows. In Sect.~\ref{sec:settings} we
discuss the experimental data and the theoretical details used in this analysis,
along with the PDF fits performed. In Sect.~\ref{sec:results}, we present the
results of these fits, we assess their quality, and we use them to understand
how the datasets and the theoretical framework affect the PDFs, in particular
in relationship with the strangeness content of the proton. We finally
provide a summary of our work in Sect.~\ref{sec:summary}.

\section{Analysis settings}
\label{sec:settings}

In this section we present the experimental datasets used as input to our
analysis, we then discuss the details of the corresponding theoretical
computations, and we finally explain which PDF fits we perform to study
their impact on the proton strangeness.

\subsection{Experimental data}
\label{sec:expdata}

The bulk of the dataset included in our analysis corresponds to the one used
in~\cite{Ball:2018iqk}, which is in turn a variant of the dataset used in the
NNPDF3.1 NNLO analysis~\cite{Ball:2017nwa}. It contains in particular
measurements of the dimuon neutrino-nucleus DIS cross sections from the
NuTeV experiment~\cite{Mason:2007zz}, and of inclusive
gauge boson production in proton-(anti)proton collisions from several Tevatron
and LHC experiments~\cite{Aaltonen:2010zza,Abazov:2007jy,Aad:2011dm,
  Aaboud:2016btc,Khachatryan:2016pev}. These
measurements represented the most constraining source of experimental
information on the strange quark and anti-quark PDFs in the NNPDF3.1 analysis.

We supplement this dataset with a number of new measurements. Concerning
neutrino-nucleus DIS, we include measurements of the ratio of dimuon to
inclusive charged-current cross sections,
$\mathcal{R}_{\mu\mu}(\omega)=\sigma_{\mu\mu}(\omega)/\sigma_{\rm CC}(\omega)$,
from the NOMAD experiment~\cite{Samoylov:2013xoa}, see Sect.~\ref{sec:theory}
for details. The data is presented for three kinematic variables $\omega$: the
neutrino beam energy $E_\nu$, the momentum fraction $x$, and the square
root of the final-state invariant mass $\sqrt{\hat{s}}$. Given that experimental
correlations are not provided amongst measurements in different kinematic
variables, only one measurement can be included in the fit at a time: we select
the $n_{\rm dat}=19$ data points as a function of $E_\nu$, the only variable
which is directly measured by the experiment among the three. We will
nevertheless verify that similar results can be obtained for instance with the
$\sqrt{\hat{s}}$--dependent dataset. The kinematic sensitivity of the NOMAD
measurements is roughly $0.03 \lsim x \lsim 0.7$, as illustrated by the coverage
of the $x$--dependent dataset.

Concerning proton-proton collisions, we augment the inclusive gauge boson
production measurement from the ATLAS experiment at a center-of-mass energy
(c.m.e.) of 7~TeV~\cite{Aaboud:2016btc} with the off-peak and forward rapidity
bins (not included in NNPDF3.1) for a total of $n_{\rm dat}=61$ data points.
Furthermore we include the $n_{\rm dat}=37$ data points corresponding to the
ATLAS (at a c.m.e. of 7~TeV)~\cite{Aad:2014xca} and CMS
(at a c.m.e. of 7~TeV and 13~TeV)~\cite{Chatrchyan:2013uja,Sirunyan:2018hde}
$W$$+$$c$ measurements; for ATLAS, we consider the charm-jet dataset, which is
amenable to fixed-order calculations (instead of the $D$-meson dataset). Finally
we take into account the $n_{\rm dat}=32$ data points corresponding to the ATLAS
$W$+jets measurement (at a c.m.e. of 8~TeV) differential in the transverse
momentum of the $W$ boson~\cite{Aaboud:2017soa}. Overall, these LHC datasets
are sensitive to the proton strangeness in the region $10^{-3}\lsim x\lsim 0.1$.
The present analysis contains a total of $n_{\rm dat}=4096$ data points;
experimental correlations within each dataset are available for all of the new
measurements considered here and are therefore included in our analysis.

\subsection{Theoretical calculations}
\label{sec:theory}

The measurements outlined in the previous section correspond to hadronic
observables already considered in~\cite{Ball:2017nwa}, except for the
ratio $\mathcal{R}_{\mu\mu}$ measured by the NOMAD experiment, and for the
production of $W$ bosons in association with light jets measured by the ATLAS
experiment. Likewise, the theoretical settings adopted in the present analysis
closely follow those described in the NNLO analysis of~\cite{Ball:2017nwa,
  Ball:2018iqk} (whereby, in particular, the charm PDF is fitted), except for
some improvements. In this section we discuss in turn the new
NOMAD observable and the theoretical details unique to the present analysis.

\subsubsection{The NOMAD ratio}

As mentioned above, the NOMAD experiment measured the ratio of dimuon to
inclusive charged-current cross sections, $\mathcal{R}_{\mu\mu}$.
Both the numerator and the denominator of $\mathcal{R}_{\mu\mu}$ are evaluated as
two-dimensional integrals of the differential cross sections over the fiducial
phase space. For the $E_\nu$--dependent data set, which we include
by default, we have
\begin{equation}
\label{eq:NOMADxsec}
\sigma_i(E_\nu)
  =
\int_{x_0}^1\frac{dx}{x}
\int_{Q^2_{\rm min}}^{Q^2_{\rm max}(x)}dQ^2
\frac{d^2\sigma_i}{dxdQ^2}(x,Q^2,E_\nu)\,,
\end{equation}
where $Q^2_{\rm max}(x)=2m_p E_{\nu}x$ and $x_0=Q^2_{\rm min}/(2m_pE_\nu)$,
with $m_p$ the proton mass. While the NOMAD measurements are reconstructed for
$Q^2\ge 1$~GeV$^2$, we assume $Q^2_{\rm min}=Q^2_0$, where $Q_0=1.65$~GeV is the
initial parametrization scale adopted in our analysis~\cite{Ball:2017nwa}.
We explicitly verified that results are unaffected if $Q^2_{\rm min}=1$~GeV$^2$
is chosen instead. The integrand in Eq.~\eqref{eq:NOMADxsec} is either the
dimuon ($i=\mu\mu$, entering the numerator or $\mathcal{R}_{\mu\mu}$) or the
inclusive ($i={\rm CC}$, entering the denominator of $\mathcal{R}_{\mu\mu}$)
charged-current cross section
\begin{align}
\frac{d^2\sigma_i}{dxdQ^2}(x,Q^2,E_\nu)
& =
\frac{G_F^2M_W^2}{4\pi}\frac{1}{(Q^2+M_W^2)^2}
\nonumber \\
& \times
\left[
\left(Y_+ - \frac{2m_p^2x^2y^2}{Q^2}\right) F_2^i(x,Q^2)
- y^2 F_L^i(x,Q^2)
+ Y_-xF_3^i
\right] K^i\,.
\label{eq:xsec}
\end{align}
The kinematic factors $Y_\pm=1\pm(1-y)^2$ are related to the inelasticity
$y=Q^2/(2m_p E_\nu x)$; $G_F$ and $M_W$ 
are respectively the Fermi constant and the mass of the $W$ boson.
The factor $K^i$ is either the identity, for $i={\rm CC}$, or the
charm semileptonic branching ratio $B_\mu$, for $i=\mu\mu$. In the latter case
we use the $E_\nu$--dependent parametrization $B_\mu(E_\nu)=a( 1 +b/E_\nu )^{-1}$,
with the values of the parameters $a$ and $b$ determined
in~\cite{Samoylov:2013xoa}, $a=0.097\pm 0.003$ and $b=6.7\pm 1.8$.
The corresponding uncertainty is included in the experimental covariance matrix
of the measurement.

\begin{figure}[!t]
\centering
\includegraphics[width=.49\textwidth]{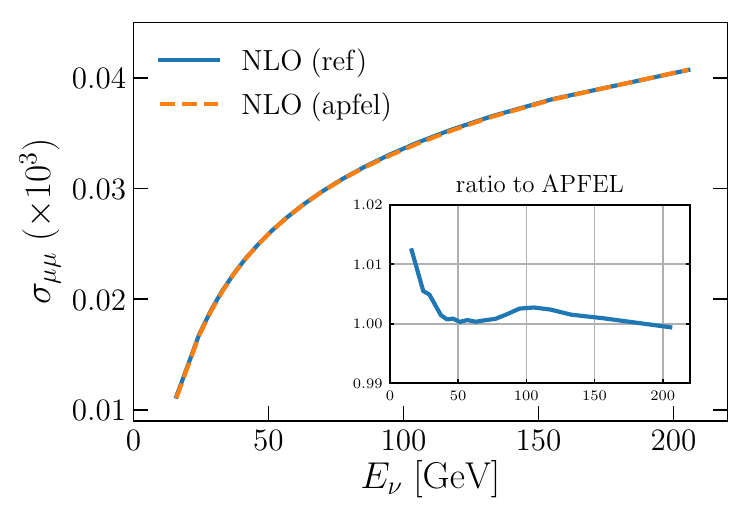}
\includegraphics[width=.49\textwidth]{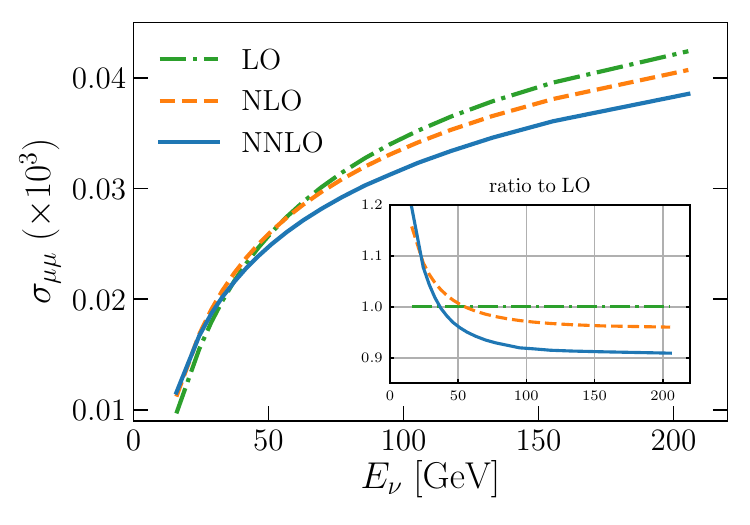}
\caption{Left: the integrated dimuon cross section as a function of the
  neutrino beam energy $E_\nu$, Eq.~\eqref{eq:NOMADxsec} (with $i=\mu\mu$),
  computed at NLO in the kinematic range measured by the NOMAD experiment
  with \textsc{ APFEL} (apfel) and with a code based on~\cite{Gao:2017kkx}
  (ref). The ratio of the two computations is shown in the inset.
  Right: the  same cross section computed in the FFN scheme ($n_f=3$) with the
  NNPDF3.1 NNLO PDF set (also the $n_f=3$ version) for various perturbative
  orders. The inset displays the ratio to the LO calculation. }
\label{fig:nnlo-kfact}
\end{figure}

Both the charm (for $i=\mu\mu$) and the total (for $i={\rm CC}$) structure
functions $F_p^i$ ($p=2,L,3$) entering Eq.~\eqref{eq:xsec} are evaluated with
\textsc{APFEL}~\cite{Bertone:2013vaa}. We benchmarked our results against
those obtained from an independent computation based on~\cite{Gao:2017kkx}.
After the correction of a bug in \textsc{APFEL}, which affected the
computation of the large-$x$ DIS coefficient functions at
next-to-leading (NLO) order, the relative difference
between the two is found to be of the order of permille, apart from
the lowest $E_{\nu}$ bins, in which it reaches the percent level,
as displayed in the left panel of Fig.~\ref{fig:nnlo-kfact}.

\subsubsection{Theoretical improvements}

In comparison to the earlier NNPDF analyses~\cite{Ball:2017nwa,Ball:2018iqk},
here we introduce several theoretical improvements, which are summarized in turn
below.

\paragraph{NNLO massive corrections in neutrino DIS.}
We incorporate the recently computed NNLO charm-quark massive
corrections~\cite{Berger:2016inr,Gao:2017kkx} in the description of the NuTeV
and NOMAD measurements. We do so by multiplying the NLO theoretical prediction
in the FONLL general-mass variable flavor number
scheme~\cite{Forte:2010ta,Ball:2011mu} by a
$K$-factor defined as the ratio between the NNLO result in the fixed-flavor
number (FFN) scheme with and without the charm-mass correction  in the matrix
  elements (ME); NNLO PDFs are used in both cases. The resulting $K$-factor
\begin{equation}
\label{eq:kfact_nnlo}
K_{\rm NNLO} \equiv \frac{\sigma_{\rm FFN}({\rm NNLO~PDFs,~NNLO~ME})}{\sigma_{\rm FFN}({\rm NNLO~PDFs,~NLO~ME)}} \, ,
\end{equation}
  is such that the prediction for the NuTeV dimuon
      cross sections becomes
\begin{equation}
   \frac{d^2\sigma_{\mu\mu}}{dxdQ^2}\Bigg|_{\rm FONLL\,(NNLO\,ME)} =
  K_{\rm NNLO} \times \frac{d^2\sigma_{\mu\mu}}{dxdQ^2}\Bigg|_{\rm FONLL\,(NLO\,ME)} \, ,
\end{equation}
 and an analogous expression holds for the NOMAD observables.

This approach
provides a good approximation of the exact result, because theoretical
predictions in the FFN and FONLL schemes are very close for the NuTeV and NOMAD
kinematics. This fact was demonstrated in~\cite{Ball:2011mu} in the case of
NuTeV data; we nevertheless checked that it remains true with the independent
computation of~\cite{Berger:2016inr,Gao:2017kkx}, and that it also applies
to the NOMAD measurements. To this purpose, we computed the relative difference
between the FONLL-A and FFN scheme predictions for the NuTeV and NOMAD datasets
based on structure functions accurate to $\mathcal{O}(\alpha_s)$. We found
that differences were less than 1\% in the entire kinematic range for NuTeV,
and of about 1.5\% irrespective of the value of $E_{\nu}$ for NOMAD. These
differences are well below the experimental and the PDF uncertainties.\footnote{
  A $\simeq 1.5\%$ difference in $K_{\rm NNLO}$ corresponds to at most a $\simeq 0.1\%$ effect
  in the absolute cross sections.}
We therefore conclude that using a NNLO $K$-factor determined in the FFN scheme in
fits that otherwise use the FONLL scheme throughout is unlikely to affect the
final results.
Given the current implementation of the
  FONLL scheme in the DIS observables~\cite{Forte:2010ta} the matching
 between the NNLO massless and massive calculations
 would require non-trivial modifications of the code of~\cite{Gao:2017kkx}, {\it e.g.}
to extract the collinear logarithms, with little practical advantage.

The $K$-factors are in general smaller than unity, and thus enhance the
(anti-)strange quark PDF when accounted for in the fit. This fact is
consistent with what was already observed in~\cite{Thorne:2019mpt}, and is
further illustrated in the right panel of Fig.~\ref{fig:nnlo-kfact}, where
we display the charm production cross section, Eq.~\eqref{eq:NOMADxsec},
with $i=\mu\mu$, as a function of $E_\nu$ in the kinematic range measured by
the NOMAD experiment. The cross section is obtained in the FFN scheme
(with $n_f=3$) at different perturbative orders using the NNPDF3.1 NNLO
PDF set (consistently with $n_f=3$). The inset displays the ratio to the
leading order (LO) calculation. Higher-order corrections clearly suppress the
cross section, in particular as $E_\nu$ increases. For instance, in the highest
energy bin the NNLO cross section is about 10\% smaller than the LO one.
The size of the NNLO correction is comparable to or larger than the size of the
NLO one, therefore its inclusion is mandatory to achieve a good description of
the data. While the comparison of Fig.~\ref{fig:nnlo-kfact} is presented in the
FFN scheme, all the fits discussed below are based on the FONLL scheme.

\paragraph{NNLO corrections in collider gauge boson production.}
Theoretical predictions for inclusive $W$- and $Z$-boson production and for
$W$-boson production in association with charm quarks or light jets are
evaluated at NLO using
\textsc{MCFM}+\textsc{APPLgrid}~\cite{Boughezal:2016wmq,Carli:2010rw}, and are
supplemented with NNLO QCD $K$-factors. These are evaluated with
\textsc{FEWZ}~\cite{Gavin:2010az} for inclusive gauge boson production, and with
the \textsc{N$_{\rm jetty}$} program~\cite{Boughezal:2015dva,Ridder:2015dxa}
for $W$-boson production with light jets. In the first case, a fixed
factorization and renormalization scale is used, equal to the mass of the gauge
boson; in the second case, a dynamic scale is used, where the factorization
and renormalization scales are defined as $\mu_F=\mu_R=\sqrt{m_W^2+p_{T,j}^2}$,
with $m_W$ the mass of the $W$ boson and $p_{T,j}$ the transverse momentum of
the hadronic jet.
Because NNLO QCD corrections 
for $W$$+$$c$ production have been presented only
  very recently~\cite{Czakon:2020coa},
in this case we accompany the data with an additional
correlated uncertainty, estimated from the 9-point scale variations of the NLO
calculation~\cite{AbdulKhalek:2019ihb,AbdulKhalek:2019bux}.

\paragraph{Nuclear corrections in neutrino DIS.}
Neutrino-DIS measurements from the NuTeV and NOMAD experiments are subject to
nuclear corrections, because they both utilize an iron (Fe) target. In this
analysis, however, we do not include such corrections because
they are expected to be subdominant
in comparison to other sources of uncertainties.
For NuTeV, they were found to be 
  moderate in a
global fit based on the same methodology used here~\cite{Ball:2018twp};
for NOMAD, they   are known to
  approximately cancel out in the ratio
$\mathcal{R}_{\mu\mu}$. To verify this last statement, we recomputed the
NOMAD ratio $\mathcal{R}_{\mu\mu}$ with the recently presented nNNPDF2.0 NLO Fe
nuclear PDF set~\cite{AbdulKhalek:2020yuc}, and compared the result with the
predictions obtained with the NLO free proton PDF set consistently determined
in~\cite{AbdulKhalek:2020yuc}. The full set of correlations between Fe and
proton PDFs were therefore appropriately taken into account.
The relative difference between the two computations (with and without nuclear
PDF corrections) turned out to range between 3\%, in the lowest $E_\nu$ bin, and
a fraction of percent, in the bins at the highest $E_\nu$. These differences are
smaller than both the data and PDF uncertainties, therefore ignoring nuclear
PDF uncertainties is a well-justified approximation.
We note that nuclear non-isoscalarity effects are treated as in~\cite{Ball:2018twp}.
 In the future, it might be interesting to extend the present analysis
  in a way that systematically accounts for nuclear PDF uncertainties.

\paragraph{Positivity of cross sections.}
We enforce the positivity of the charm structure function $F_2^c$ with a
procedure similar to that described in~\cite{Ball:2014uwa} for light quarks.
This additional positivity constraint is required to prevent the fitted charm
PDF becoming unphysically negative once the new datasets are included in the
fit.

\subsection{PDF sets}
\label{sec:PDFsets}

We assess the impact of the datasets and of the theoretical choices outlined
in Sects.~\ref{sec:expdata}-\ref{sec:theory} on PDFs by performing the series of
fits summarized in Table~\ref{table:fitoverview}. All of them are accurate to
NNLO in perturbative QCD (where available), and are based on the
NNPDF methodology, see~\cite{Ball:2014uwa} and references therein for a
comprehensive description.

The first fit ({\tt str\_base}) is our baseline, and corresponds to the fit
of~\cite{Ball:2018iqk} with the addition of the NNLO charm-mass $K$-factors
for the NuTeV data and of the positivity constraint on $F_2^c$,
 and with the removal of the 2010 and 2011 ATLAS $W,Z$ inclusive
measurements of~\cite{Aad:2012sb,Aaboud:2016btc}. We will present
a comparison of this baseline fit with the NNPDF3.1 PDF set
of~\cite{Ball:2017nwa} in Sect.~\ref{sec:PDFs}. This fit is then
supplemented with all the new LHC data, 
  including the ATLAS $W,Z$
  measurements from~\cite{Aad:2012sb,Aaboud:2016btc}, to obtain the second fit
({\tt str\_prior}), for which we generate $N_{\rm rep}=850$ Monte Carlo replicas.
 The exclusion of the ATLAS $W,Z$ measurements
  of~\cite{Aad:2012sb,Aaboud:2016btc} from the {\tt str\_base} fit allows us
  to quantify the effect of the full LHC strangeness-sensitive
  dataset by comparing this fit with the {\tt str\_prior} one.
This second fit, {\tt str\_prior}, is finally further supplemented
with the NOMAD data,
specifically the set that depends on $E_\nu$, to determine the third fit
({\tt str}). Bayesian reweighting and unweighting~\cite{Ball:2011gg,Ball:2010gb}
are used in this last step, because they allow one to evaluate the
two-dimensional integral in Eq.~\eqref{eq:NOMADxsec} only once, a task that
would otherwise be computationally very intensive in a fit. After reweighting,
one ends up with $N_{\rm eff}=105$ effective replicas, from which we
construct a set of $N_{\rm rep}=100$ replicas.

\begin{table}[!t]
  \scriptsize
  \renewcommand{\arraystretch}{1.30}
  \begin{tabularx}{\linewidth}{XXX}
  \toprule
       Fit ID
     & Dataset
     & Theory
     \\
    \hline
     \multirow{3}{*}{\tt str\_base}
     & \multirow{2}{*}{same as~\cite{Ball:2018iqk}  $-$}
     & same as~\cite{Ball:2018iqk} +
     \\
     &   \multirow{2}{*}{ATLAS
        $W$, $Z$~\cite{Aad:2012sb,Aaboud:2016btc}}
     & NNLO $K$-fact NuTeV
     \\
     &
     & $F_2^c$ positivity
     \\
    \hline
     \multirow{5}{*}{{\tt str\_prior}}
     & same as {\tt str\_base} +
     & same as {\tt str\_base} +
     \\
     & ATLAS $W$, $Z$ (full)~\cite{Aad:2012sb,Aaboud:2016btc}
     & correlated theory
     \\
     & ATLAS $W$$+$$c$~\cite{Aad:2014xca} 
     & uncertainty for
     \\
     & CMS $W$$+$$c$~\cite{Chatrchyan:2013uja,Sirunyan:2018hde} 
     & unknown NNLO QCD 
     \\  
     & ATLAS $W$+jets~\cite{Aaboud:2017soa}
     & corrections
     \\
    \hline
     \multirow{2}{*}{{\tt str}}
     & same as {\tt str\_prior} +
     & same as {\tt str\_prior} +
     \\  
     & NOMAD $E_{\nu}$ set~\cite{Samoylov:2013xoa}
     & NNLO $K$-fact NOMAD
     \\
     \hline
     \multirow{2}{*}{{\tt str\_s\_hat}}
     & same as {\tt str\_prior} +
     & \multirow{2}{*}{same as {\tt str}}
     \\  
     & NOMAD $\sqrt{\hat{s}}$ set~\cite{Samoylov:2013xoa}
     &
     \\
      \hline
       \multirow{2}{*}{{\tt str\_prior\_pch}}
     & \multirow{2}{*}{same as {\tt str\_prior} }
     & same as {\tt str\_prior},
     \\  
     & 
     & perturbative charm
     \\
     \hline
       \multirow{2}{*}{{\tt str\_pch}}
     & same as {\tt str\_prior\_pch} +
     & same as {\tt str\_prior\_pch} +
     \\  
     & NOMAD $E_{\nu}$ set~\cite{Samoylov:2013xoa}
     & NNLO $K$-fact NOMAD
     \\
    \bottomrule
  \end{tabularx}
  \vspace{0.3cm}
  \caption{A list of the PDF fits presented in this work, see text for
    details.}
  \label{table:fitoverview}
\end{table}

We also produced variants of these three fits. First of all, in order to assess the
impact of the choice of the specific NOMAD dataset, we performed the
{\tt str\_s\_hat} fit, which is equivalent to the {\tt str} fit, except for the
fact that the NOMAD $E_\nu$--dependent dataset is replaced by its
$\sqrt{\hat{s}}$--dependent counterpart. In this case, after reweighting one
ends up with $N_{\rm eff}=135$ effective replicas (out of $N_{\rm rep}=850$ initial
replicas), from which we construct an ensemble of $N_{\rm rep}=100$ replicas.
Second, in order to assess the impact of parametrizing the charm PDF, we performed the
{\tt str\_prior\_pch} and {\tt str\_pch} fits. These fits are equivalent to the
{\tt str\_prior} and {\tt str} fits, except for the fact that the charm PDF is
generated perturbatively off the gluon and the light quark PDFs. In this case, we
produced only $N_{\rm rep}=500$ replicas in the {\tt str\_prior\_pch} fit; after
reweighting we are left with $N_{\rm eff}=157$ effective replicas, from which we
constructed an ensemble of $N_{\rm rep}=100$ replicas for the {\tt str\_pch} fit.

\section{Results}
\label{sec:results}

In this section we present the main results of our analysis. First, we discuss the
quality of the fits that we performed; then, we compare the data to our 
theoretical predictions; next, we present the PDFs that we
determine; and finally, we revisit the strangeness content of the proton in
light of them. We conclude by focusing on the impact of the NOMAD dataset 
and of the implications that the treatment of the charm PDF has on
our results.

\subsection{Fit quality}
\label{sec:fitquality}

In Table~\ref{table:chi2} we summarize the values of the $\chi^2$ per data
point obtained from five of the six fits discussed in Sect.~\ref{sec:PDFsets},
see also Table~\ref{table:fitoverview}: $\chi^2_{\rm base}$ for {\tt str\_base};
$\chi^2_{\rm pr}$ for {\tt str\_prior}; $\chi^2_{\rm str}$ for {\tt str};
$\chi^2_{\rm str\_s\_hat}$ for {\tt str\_s\_hat}; and $\chi^2_{\rm str\_pch}$ for
{\tt str\_pch}. The $\chi^2$ of the {\tt str\_prior\_pch} fit is not reported
because it is not particularly more informative than the one of the
{\tt str\_pch} fit, which includes the complete dataset. In all cases, the
value of the $\chi^2$ per data point correspond to the definition given in
Eqs.~(3.2)-(3.3) in~\cite{Nocera:2019wyk}. The values in square brackets are
for datasets not included in the corresponding fit.

\begin{table}[!t]
  \scriptsize
  \renewcommand{\arraystretch}{1.30}
  \begin{tabularx}{\linewidth}{XXcccccc}
    \toprule
       Process
     & Dataset
     & $n_{\rm dat}$
     & $\chi^2_{\rm base}$
     & $\chi^2_{\rm pr}$
     & $\chi^2_{\rm str}$
     & $\chi^2_{\rm str\_s\_hat}$
     & $\chi^2_{\rm str\_pch}$   \\
 \hline
 $\nu$DIS ($\mu\mu$) &                                                & 76/76/95/91/95           &
  0.70  & 0.71  & 0.53 & 0.52 & 0.63 \\
   & NuTeV~\cite{Mason:2007zz}                      & 76/76/76/76/76           &
   {  0.70}  & 0.71  & 0.53 & 0.55 & 0.61 \\
                     & NOMAD~\cite{Samoylov:2013xoa}                  & ---/---/19/15/19         & {  [9.0]}  & [8.8] & 0.55 & 0.35 & 0.69 \\
 \hline
 $W$, $Z$ (incl.)    &
 & { 327}/418/418/418/418      &
   {  1.38}  & 1.40  & 1.40 & 1.39 & 1.40 \\
 & ATLAS~\cite{Aaboud:2016btc}
 & { ---}/61/61/61/61           &
   {  3.22}  & 1.65  & 1.67 & 1.64 & 1.80 \\
 \hline
 $W$$+$$c$           &                                                &  ---/37/37/37/37         & {  [0.76]} & 0.68  & 0.60 & 0.66 & 0.68 \\
                     & CMS~\cite{Chatrchyan:2013uja,Sirunyan:2018hde} &  ---/15/15/15/15         & {  [1.10]} & 0.98  & 0.96 & 1.00 & 1.00 \\
                     & ATLAS~\cite{Aad:2014xca}                       &  ---/22/22/22/22         & {  [0.53]} & 0.48  & 0.42 & 0.43 & 0.46 \\
 \hline
 $W$+jets            & ATLAS~\cite{Aaboud:2017soa}                    &  ---/32/32/32/32         & {  [1.58]} & 1.18  & 1.18 & 1.18 & 1.18 \\
 \hline
     {\bf Total}         &                                                & { 3917}/4077/4096/4092/4096 &
     {  1.17}  & 1.17  & 1.17 & 1.17 & 1.20 \\
 \bottomrule
  \end{tabularx}
  \vspace{0.3cm}
  \caption{Values of the $\chi^2$ per data point for the
    strangeness-sensitive datasets discussed in this work obtained from
    the {\tt str\_base}, {\tt str\_prior}, {\tt str}, {\tt str\_s\_hat},
    and {\tt str\_pch} fits, see Table~\ref{table:fitoverview}. Values in
    square brackets are for datasets not included in the corresponding fit.}
  \label{table:chi2}
\end{table}

We first assess the general consistency of the new experimental data, by comparing the values
of the $\chi^2$ of the first three fits. The description of the new datasets
--- which, in particular, is not optimal for the ATLAS $W$, $Z$ dataset in the
{\tt str\_base} fit and for the NOMAD dataset in the {\tt str\_base} and
{\tt str\_prior} fits --- markedly improves as soon as they are included in
subsequent fits. The largest effect is witnessed by the NOMAD dataset, whose
$\chi^2$  per data point decreases from about 9 in the {\tt str\_base} and {\tt str\_prior}
fits to about 0.6 in the {\tt str} fit. The $\chi^2$ for all of the other
datasets is in general not affected upon the addition of the NOMAD dataset in
the {\tt str} fit, except for the NuTeV dataset, whose $\chi^2$ is further
improved in comparison to the {\tt str\_prior} fit. We therefore conclude that
the global dataset is overall consistent and satisfactorily described in the final
{\tt str} fit.

We then assess the consistency of alternative NOMAD datasets by comparing
the $\chi^2$ of the {\tt str} and {\tt str\_s\_hat} fits. We recall that they
include, respectively, the $E_\nu$--dependent and the
$\sqrt{\hat{s}}$--dependent distributions. A very similar fit quality is 
achieved in the two cases, not only for the NOMAD dataset, but also for all of
the other datasets: the differences in the values of the $\chi^2$ between the
two fits are smaller than statistical fluctuations. This fact suggests that the
alternative NOMAD datasets are consistent between them and with the rest of the
dataset. This conclusion is in line with the observation that a similar number
of effective replicas is obtained by reweighting the {\tt str\_prior} fit
with either dataset, see the discussion in Sect.~\ref{sec:PDFsets}.

We finally assess the effect of parametrizing the charm PDF (or not)
by comparing the $\chi^2$ of the {\tt str} and {\tt str\_pch} fits. We recall
that the two fits contain exactly the same datasets, however in the former the
charm PDF is parametrized on the same footing as the other light quark PDFs,
while in the latter it is generated perturbatively off the light quarks and
the gluon. The fitted charm fit ({\tt str}) achieves a better description of
the strangeness-sensitive datasets, and of the global dataset overall, than the
perturbative charm fit ({\tt str\_pch}). We note in particular the $\chi^2$
values of the ATLAS $W,Z$ and of the total datasets, which increase
respectively from 1.67 to 1.80 and from 1.17 to 1.20 when comparing the
{\tt str} and the {\tt str\_pch} fits. We therefore confirm previous
studies indicating that fitting the charm PDFs
improves the description of the experimental data within a global
PDF analysis.

\subsection{Comparison with experimental data}

We now compare the strangeness-sensitive datasets included in our analysis
with the corresponding theoretical predictions. Our aim is to assess the
impact of the various datasets. To this purpose, we compare the fits obtained
without and with a specific dataset included.

\begin{figure}[!t]
\centering 
\includegraphics[width=.49\textwidth,clip=true trim=0 0.4cm 0 0]{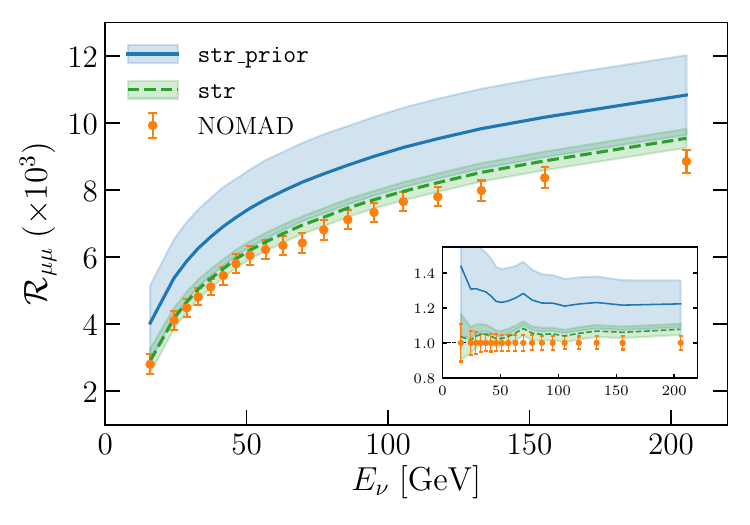}
\includegraphics[width=.49\textwidth,clip=true trim=0 0.4cm 0 0]{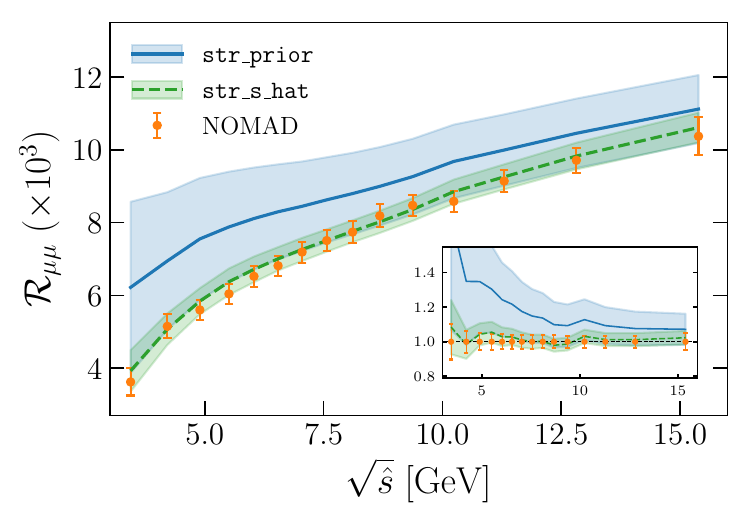} 
\caption{Comparison between the theoretical predictions, obtained from the
  {\tt str\_prior} and {\tt str} (left) or {\tt str\_s\_hat} (right) fits,
  and the experimental data for the $E_{\nu}$-- and $\sqrt{\hat{s}}$--dependent
  NOMAD measurements. The insets display the ratio to the central value of
  each data point. {  The error bands on the theory predictions
    indicate the one-sigma PDF uncertainties.}}
\label{fig:nomad_data_vs_th_Enu}
\end{figure}

We first focus on the neutrino-DIS datasets. In
Fig.~\ref{fig:nomad_data_vs_th_Enu} we display the comparison for the
$E_\nu$--dependent and the $\sqrt{\hat{s}}$--dependent NOMAD measurements.
We compare the theoretical predictions obtained from the {\tt str\_prior} fit
and, respectively, either from the {\tt str} or the {\tt str\_s\_hat} fits.
The insets display the ratio to the central value of each measured data point.
In the two cases, we observe a consistent picture: the theoretical prediction
obtained from the {\tt str\_prior} fit overshoots the data points by about
20\% (10\%) for the $E_\nu$--dependent ($\sqrt{\hat{s}}$--dependent) dataset;
after reweighting, the theoretical prediction nicely describes the data points
with an uncertainty consistently reduced by up to a factor of four. We
explicitly checked that the same reduction of the uncertainty occurs also in
the case of perturbative charm fits without ({\tt str\_prior\_pch}) and with
({\tt str\_pch}) the $E_\nu$--dependent NOMAD dataset included. In this case,
however, the underlying PDFs and the strangeness ratio $R_s$ vary in comparison
to fitted charm fits, as discussed in Sect.~\ref{sec:strange}.

\begin{figure}[!t]
  \centering
  \includegraphics[width=.49\textwidth,clip=true,trim=0 0.4cm 0 0]{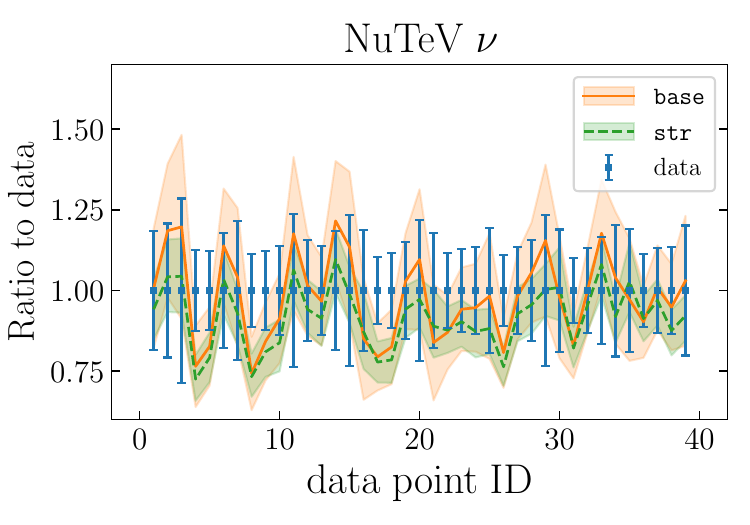}
  \includegraphics[width=.49\textwidth,clip=true,trim=0 0.4cm 0 0]{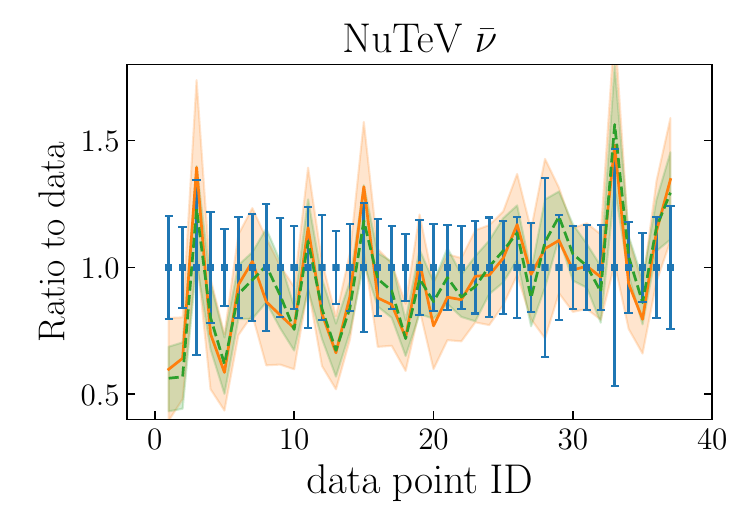}
  \caption{Comparison between theoretical predictions and experimental data for
  the neutrino (left) and antineutrino (right) charm dimuon cross sections
  measured by the NuTeV experiment~\cite{Mason:2007zz}. Data and theory are
  normalized to the central value of the former; data points are sorted by their
  ID values, roughly corresponding to increasing $x$ and $Q^2$ values (for fixed
  pseudo-rapidity bins $y$) when the plots are read from left to right.}
  \label{fig:NuTeV}
\end{figure}

In Fig.~\ref{fig:NuTeV} we display the data/theory comparison for the charm
dimuon cross sections from the NuTeV measurement of~\cite{Mason:2007zz}
(for both neutrino and antineutrino beams). In this case, predictions are
determined from the {\tt str\_base} and {\tt str} PDF input sets, and are
normalized to the central value of the data points. These are sorted by their
ID value, roughly corresponding to increasing $x$ and $Q^2$ values
(for fixed pseudo-rapidity bins $y$) when the plot is read from left to right.
A fair agreement between data and theory
is observed, as expected from the pattern of the $\chi^2$ values reported in
Table~\ref{table:chi2}. The inclusion of the NOMAD data in the {\tt str} fit
suppresses the theoretical expectation for the NuTeV neutrino cross sections
(but not for the antineutrino ones, for which no analogue observable 
is measured by NOMAD);
uncertainties are reduced by up to a factor of two (again, more markedly for
the neutrino data points than for the antineutrino ones). Both the shift in the
central value and the reduction of the uncertainty remain smaller than the
comparatively large experimental uncertainty.

\begin{figure}[!t]
  \centering
  \includegraphics[width=.49\textwidth,clip=true,trim=0 0.4cm 0 0]{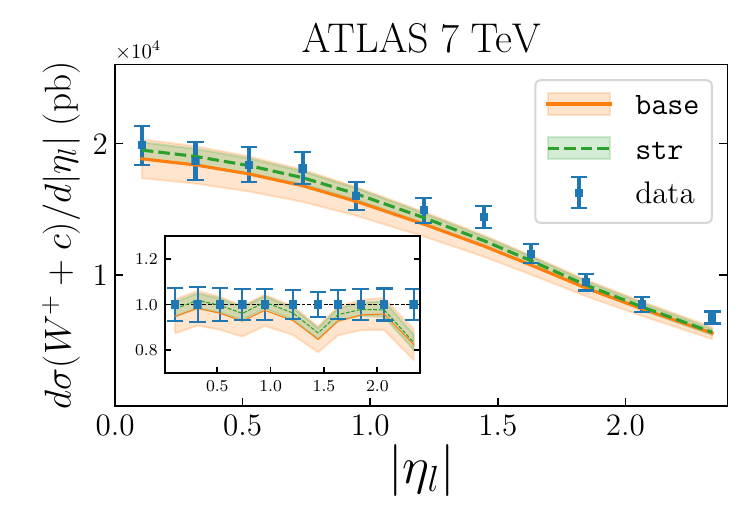}
  \includegraphics[width=.49\textwidth,clip=true,trim=0 0.4cm 0 0]{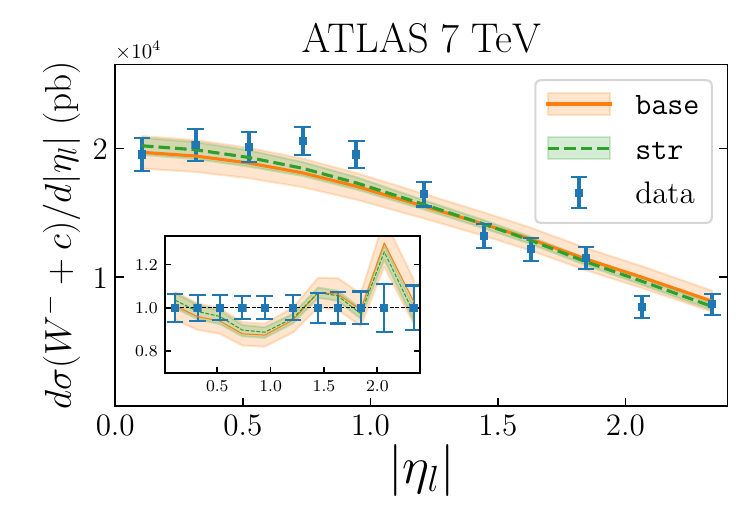}
  \includegraphics[width=.49\textwidth,clip=true,trim=0 0.4cm 0 0]{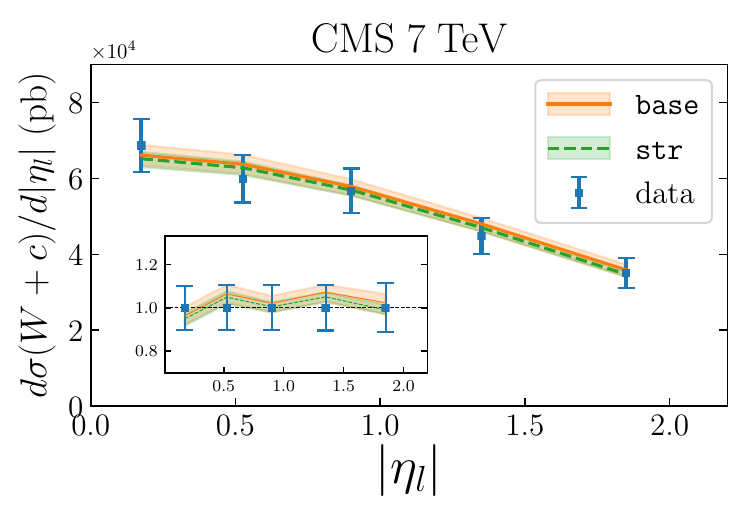}
  \includegraphics[width=.49\textwidth,clip=true,trim=0 0.4cm 0 0]{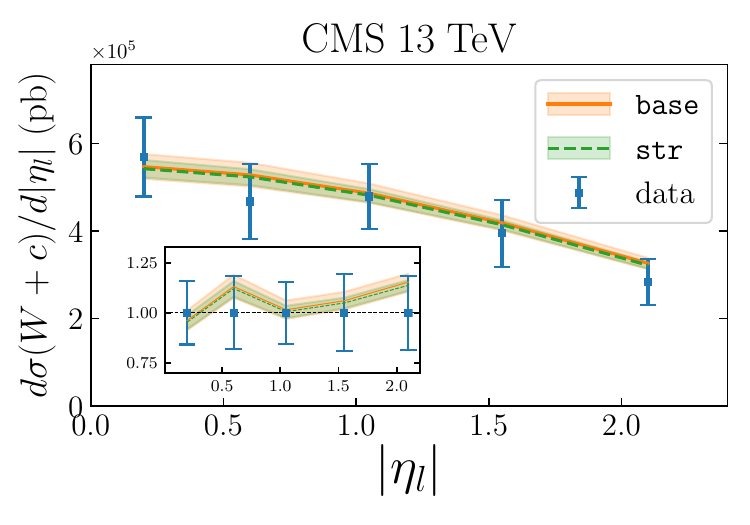}
  \includegraphics[width=.49\textwidth,clip=true,trim=0 0.4cm 0 0]{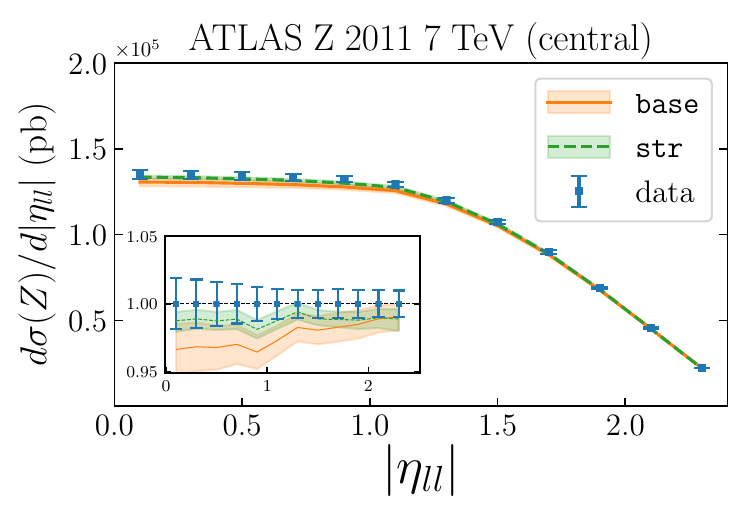}
  \includegraphics[width=.49\textwidth,clip=true,trim=0 0.4cm 0 0]{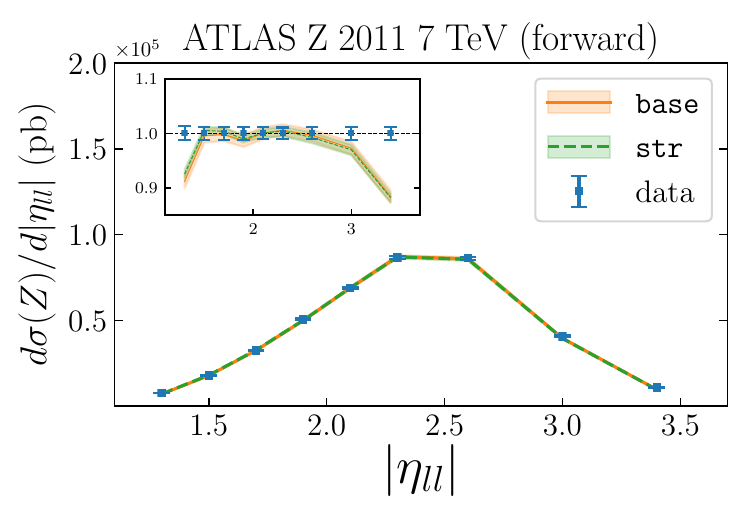}
\caption{Comparison between theoretical predictions and experimental data for
  some of strangeness-sensitive proton collider measurements used in this work.
  Top: the $W$$+$$c$ lepton rapidity distributions (separately for $W^+$ and
  $W^-$) corresponding to the ATLAS measurement at 7 TeV~\cite{Aad:2014xca}.
  Middle: the $W$$+$$c$ lepton rapidity distributions (sum of $W^+$ and $W^-$)
  corresponding to the CMS measurements at 7 TeV~\cite{Chatrchyan:2013uja} and
  13 TeV~\cite{Sirunyan:2018hde}. Bottom: the $Z$ dilepton rapidity
  distributions from the ATLAS measurement of~\cite{Aaboud:2016btc}. The insets
  display the theory to data ratio. Theoretical predictions are evaluated with
  the {\tt str\_base} and {\tt str} fits.}
\label{fig:atlas_wcharm_wp_7tev_data_vs_th}
\end{figure}

We now turn to the hadron collider data.
In Fig.~\ref{fig:atlas_wcharm_wp_7tev_data_vs_th}
we display: the $W$$+$$c$ lepton rapidity distributions corresponding to the
ATLAS measurement of~\cite{Aad:2014xca} (for both $W^+$ and $W^-$) and to
the CMS measurements (sum of $W^+$ and $W^-$)
of~\cite{Chatrchyan:2013uja,Sirunyan:2018hde} (respectively at a c.m.e.~of
7~TeV and 13~TeV); and the $Z$ dilepton rapidity distributions from the ATLAS
measurement of~\cite{Aaboud:2016btc} at a c.m.e.~of 7 TeV 
(for both the central and forward
selection cuts). The insets display the ratio of the theory to the central
value of the experimental measurement. As in Fig.~\ref{fig:NuTeV}, theoretical
predictions are evaluated with the {\tt str\_base} and {\tt str} PDF sets.

A fair agreement between data and theory is found in all cases, as expected
from the pattern of $\chi^2$ values reported in Table~\ref{table:chi2}.
However, we clearly see that the size of the PDF uncertainty relative to the
size of the data uncertainty depend on the dataset. Concerning the
ATLAS and CMS $W$$+$$c$ measurements, experimental uncertainties span the range
between 10\% and 20\%, and are consistently larger than PDF uncertainties.
We note that the PDF uncertainties in the theory predictions are markedly
reduced in the {\tt str} fit in comparison to {\tt str\_base} fit,
as highlighted by the ratios in the insets.
Concerning the ATLAS $Z$ distribution, the total experimental uncertainty is
much smaller than the $W$ counterpart, around 2\% for the central
rapidity bin, and in the central region it is comparable to the
PDF uncertainty. We therefore expect this measurement to be one of the most
constraining amongst all of the LHC measurements considered in this work.
Interestingly, once the NOMAD dataset is included in the fit, the central value
of the theoretical prediction approaches the central value of the ATLAS data,
and PDF uncertainties are slightly reduced. A similar trend can be observed for
the forward selection data. This behaviour is a
further sign of the good overall compatibility of all of the datasets, and
in particular of neutrino DIS and LHC gauge boson production measurements.

\subsection{Parton distributions}
\label{sec:PDFs}

We now turn to study the impact of the theoretical assumptions and of the new
datasets considered in this analysis on the PDFs. We first present
a comparison between the {\tt str\_base} and the NNPDF3.1 parton sets, and
then a comparison amongst the {\tt str\_base}, {\tt str\_prior} and {\tt str}
PDF sets. In the latter case, because the new datasets are expected to mainly
affect the strange quark and anti-quark distributions, we will
focus on the total and valence strange distributions, first, and on the other
PDFs, then.
 
\subsubsection{Comparison with NNPDF3.1}

Our baseline fit {\tt str\_base} differs from NNPDF3.1~\cite{Ball:2017nwa} in
several respects. As explained in Sect.~\ref{sec:theory}, these include: the
treatment of inclusive jet production from ATLAS and CMS with NNLO $K$-factors,
see~\cite{Ball:2018iqk}; an updated treatment of non-isoscalarity effects
in neutrino-DIS data, see~\cite{Ball:2018twp}; the
inclusion of the NNLO massive corrections to the NuTeV structure functions;
the new $F_2^c$ positivity constraint; and the correction of the {\sc APFEL}
bug found in the benchmark reported in Fig.~\ref{fig:nnlo-kfact},
which affected the large-$x$ implementation of the NLO coefficient
functions.
{  Furthermore following the motivation
  presented in Sect.~\ref{sec:PDFsets}, the ATLAS $W,Z$ rapidity distributions
from~\cite{Aad:2012sb,Aaboud:2016btc} that were part of NNPDF3.1 are excluded from {\tt str\_base}.}

\begin{figure}[!p]
 \centering
 \includegraphics[width=.99\textwidth]{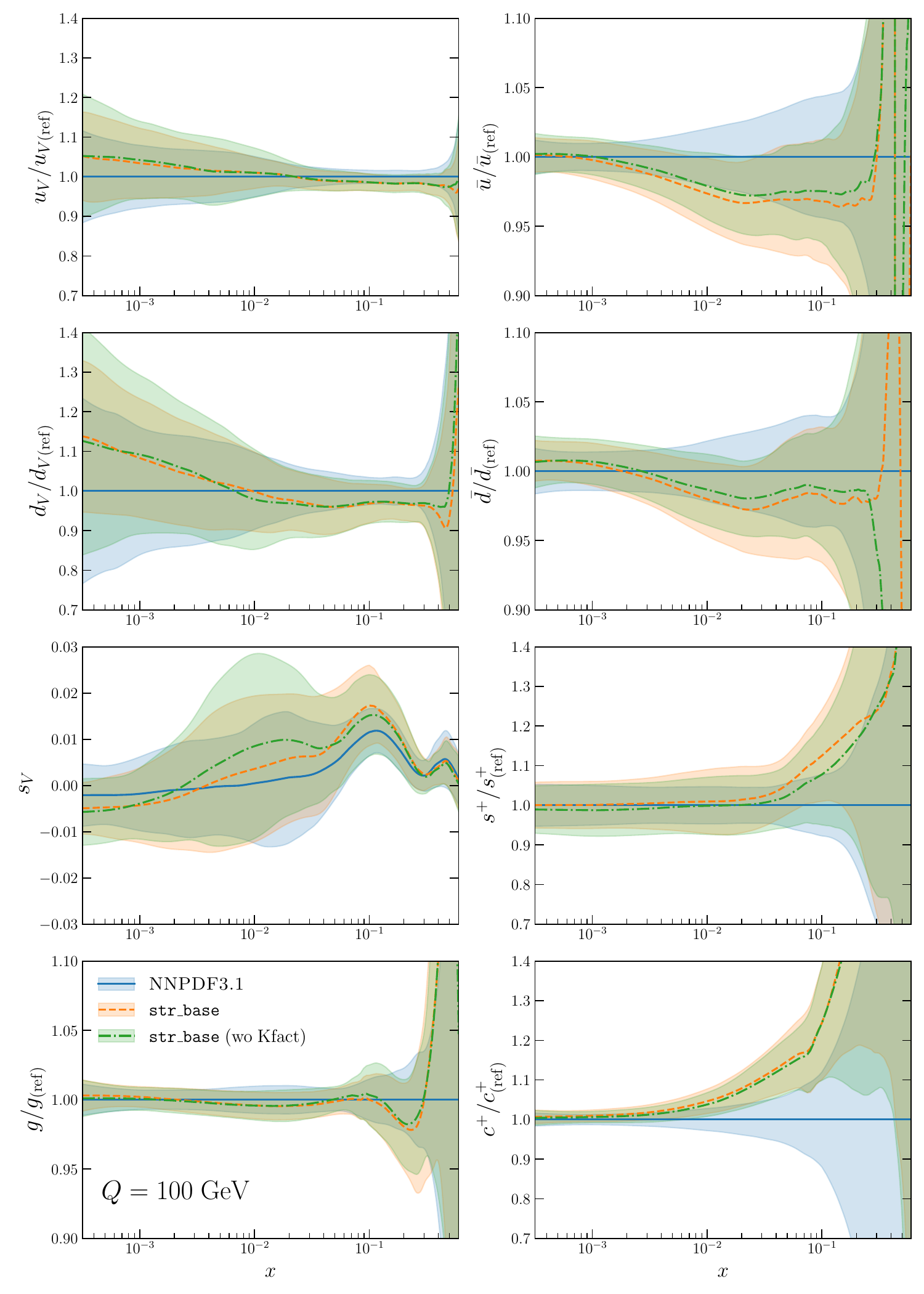}
 \caption{Comparison between the NNPDF3.1 NNLO fit~\cite{Ball:2017nwa}
   and the baseline fit used in this work, {\tt str\_base}. From top to bottom
   and left to right we show the up (valence and sea), down (valence and sea),
   strange (valence and total), gluon, and charm distributions at a scale
   $Q=100$~GeV; PDFs are normalized to the central value of the NNPDF3.1 parton
   set (ref), except for the strange valence distribution, for which the
   absolute PDFs are shown.}
\label{fig:nnpdf31comp}
\end{figure}

In order to gauge the impact of all these differences, in
Fig.~\ref{fig:nnpdf31comp} we compare the NNPDF3.1 and {\tt str\_base} parton
sets. We display the up (valence and sea), down (valence and sea), strange
(valence and total), gluon, and charm distributions at a scale $Q=100$~GeV; PDFs
are normalized to the central value of the NNPDF3.1 parton set, except for the
strange valence distribution, for which the absolute PDFs are shown.
{  In addition, we  also display in this comparison
  the results of a variant of {\tt str\_base}
  obtained without the NNLO $K$-factors, Eq.~(\ref{eq:kfact_nnlo}),
  for the NuTeV cross sections, in order to isolate their impact in the resulting PDFs.}

In comparison to NNPDF3.1, in the {\tt str\_base} fit we
observe: a rearrangement of the quark flavor separation at medium and large $x$;
an increase in the central value of the strange PDF for
$x\gsim 10^{-3}$; a
similar effect in the case of the charm PDF for $x\gsim 10^{-2}$ (mostly
due to the new $F_2^c$ positivity constraint); and a harder gluon at
large $x$ (mostly due to the improved NNLO treatment of jet data).
All in all, while the two fits agree within uncertainties, the improvements
introduced in the {\tt str\_base} fit lead to PDF differences that are
sufficiently large to adopt it as baseline in the current study. Therefore
the NNPDF3.1 set will not be discussed further in the sequel.

{  From the comparison of {\tt str\_base} with its variant
  without the $K$-factors, the most noticeable effect
  is the moderate suppression of $s^+$ in the
  region $x\gsim 0.05$, which can represent a shift of  up to half a sigma
  in units of the PDF uncertainty.
 One can also observe a small correlated
  enhancement of the up and down quark sea distributions in the
  same region of $x$. The impact of the  $K$-factors turns out to be
  negligible for the gluon and other flavour combinations.}

\subsubsection{Total and valence strange distributions}

\begin{figure}[!t]
\includegraphics[width=.49\textwidth]{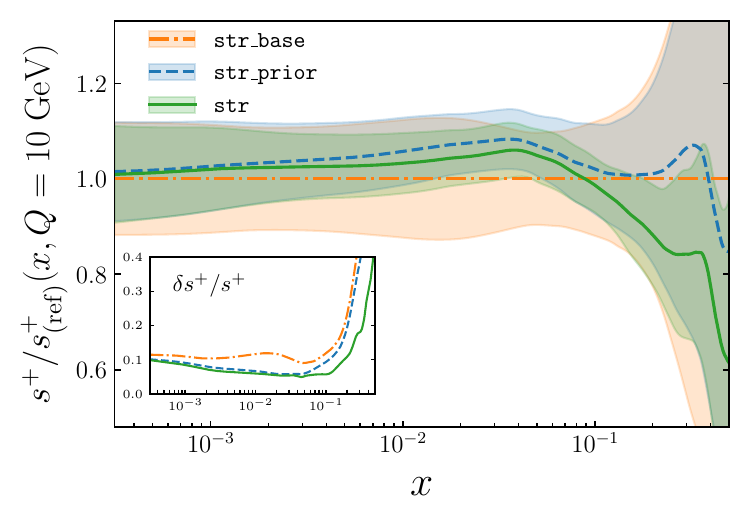}
\includegraphics[width=.49\textwidth]{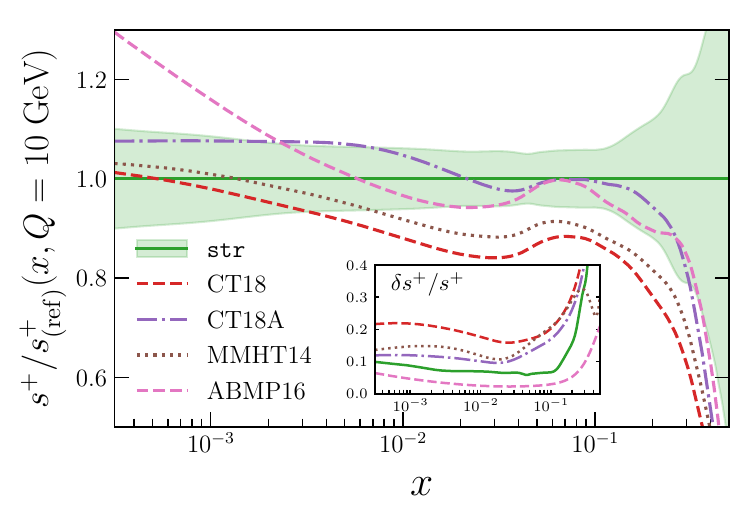}\\
\includegraphics[width=.49\textwidth,clip=true,trim= 0.35cm 0.5cm 0 0]{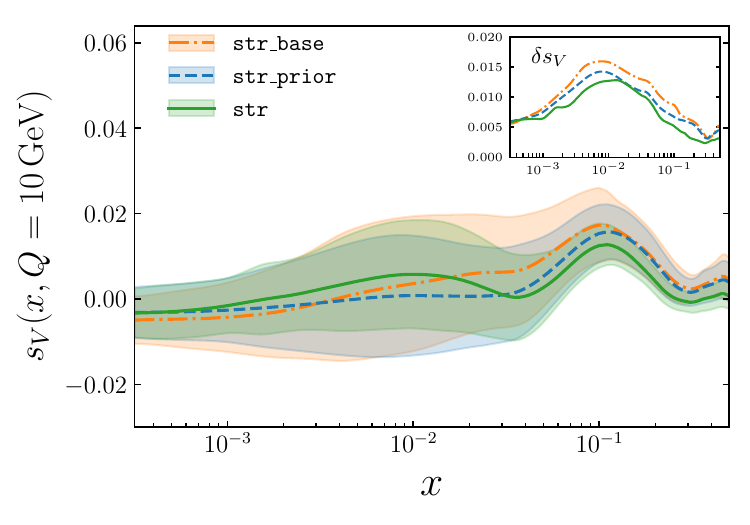}
\includegraphics[width=.49\textwidth,clip=true,trim= 0.35cm 0.5cm 0 0]{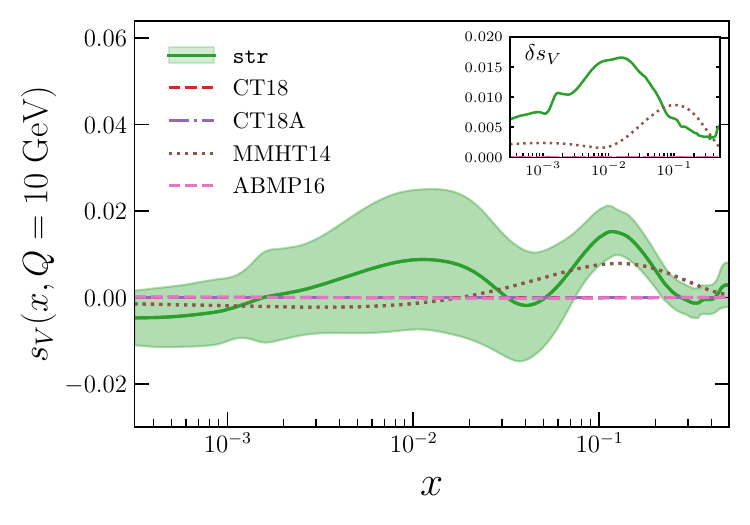}\\
\caption{A comparison of the total (top) and valence (bottom) strange
  distributions, $s^+$ and $s_V$, at $Q=10$~GeV. PDFs are from the
  {\tt str\_base}, {\tt str\_prior} and {\tt str} fits (left panels) and from
  the {\tt str} and other recent PDF fits (right panels), see text for details.
  The total strangeness $s^+$ is normalized to the central value of the
  {\tt str\_base} fit. The insets display the corresponding relative
  ($\delta s^+/s^+$) and absolute ($\delta s_V$) PDF uncertainties.}
\label{fig:strange}
\end{figure}

In Fig.~\ref{fig:strange} we display the total and valence strange
distributions
at $Q=10$~GeV.
We compare in turn the PDFs obtained from the {\tt str\_base}, {\tt str\_prior}
and {\tt str} fits, and those obtained from the {\tt str} fit with other
recent parton sets. Specifically, we consider CT18, CT18A~\cite{Hou:2019efy}
(CT18A is a variant of CT18 that includes the ATLAS $W$, $Z$ data),
MMHT14~\cite{Harland-Lang:2014zoa}, and ABMP16~\cite{Alekhin:2017kpj}.
They all include only a subset of the strangeness-sensitive data included in
our analysis (see Table~\ref{table:chi2}), in particular: the NuTeV dataset is
part of all PDF sets; the NOMAD dataset is only part of ABMP16; and
the off-peak and forward ATLAS $W$, $Z$ bins, the $W$$+$$c$ and the $W$+jets
datasets are not part of any of these PDF sets.
We also emphasize that, apart from the more extensive dataset, our analysis
differs from all of the other PDF determinations shown in
Fig.~\ref{fig:strange} in that the charm-quark PDF is
fitted on the same footing as the other light-quark PDFs~\cite{Ball:2016neh}.
This feature was demonstrated to improve the description
of DIS and LHC datasets, and in particular to partially relieve tensions
between the NuTeV and the ATLAS $W$, $Z$ datasets~\cite{Ball:2017nwa}.
The insets in Fig.~\ref{fig:strange} display the relative and absolute PDF
uncertainties for the total ($\delta s^+/s^+$) and valence ($\delta s_V$)
strange distributions, respectively. In the case of $s^+$, the curves
are normalized to the central value of the {\tt str\_base} fit.

A comparison amongst the {\tt str\_base}, {\tt str\_prior} and {\tt str} fits
reveals that the impact of the data is consistent for the total and valence
strange distributions. The inclusion of the LHC datasets in the
{\tt str\_prior} fit does not alter the central value of the PDFs in a
significant way, while it narrows the PDF uncertainty across most of the
$x$ range. The inclusion of the NOMAD dataset in the {\tt str} fit is
associated to a larger effect: the central value of both $s^+$ and $s_V$ is
suppressed by about 20\% (or more) for $x\gtrsim 0.1$; the uncertainty is
further reduced by up to a third in the same $x$ region.

A comparison amongst the {\tt str} fit and other recent parton sets reveals
differences in the shape of the central value of the $s^+$ and $s_V$
distributions. In the second case, in particular, only
MMHT14~\cite{Harland-Lang:2014zoa} allows for a non-zero parametrization.
Within the larger uncertainties of the CT18, CT18A and MMHT14 PDF sets, however,
results are overall consistent. In this respect, note that
the very small uncertainty of the ABM16 set is an artifact of the lack of a
tolerance criterion in their analysis. In this case, uncertainties should be
rescaled by a factor which is however not determined in their analysis.

\subsubsection{Light quark, charm, and gluon PDFs}

\begin{figure}[!t]
 \centering
 \includegraphics[width=.90\textwidth]{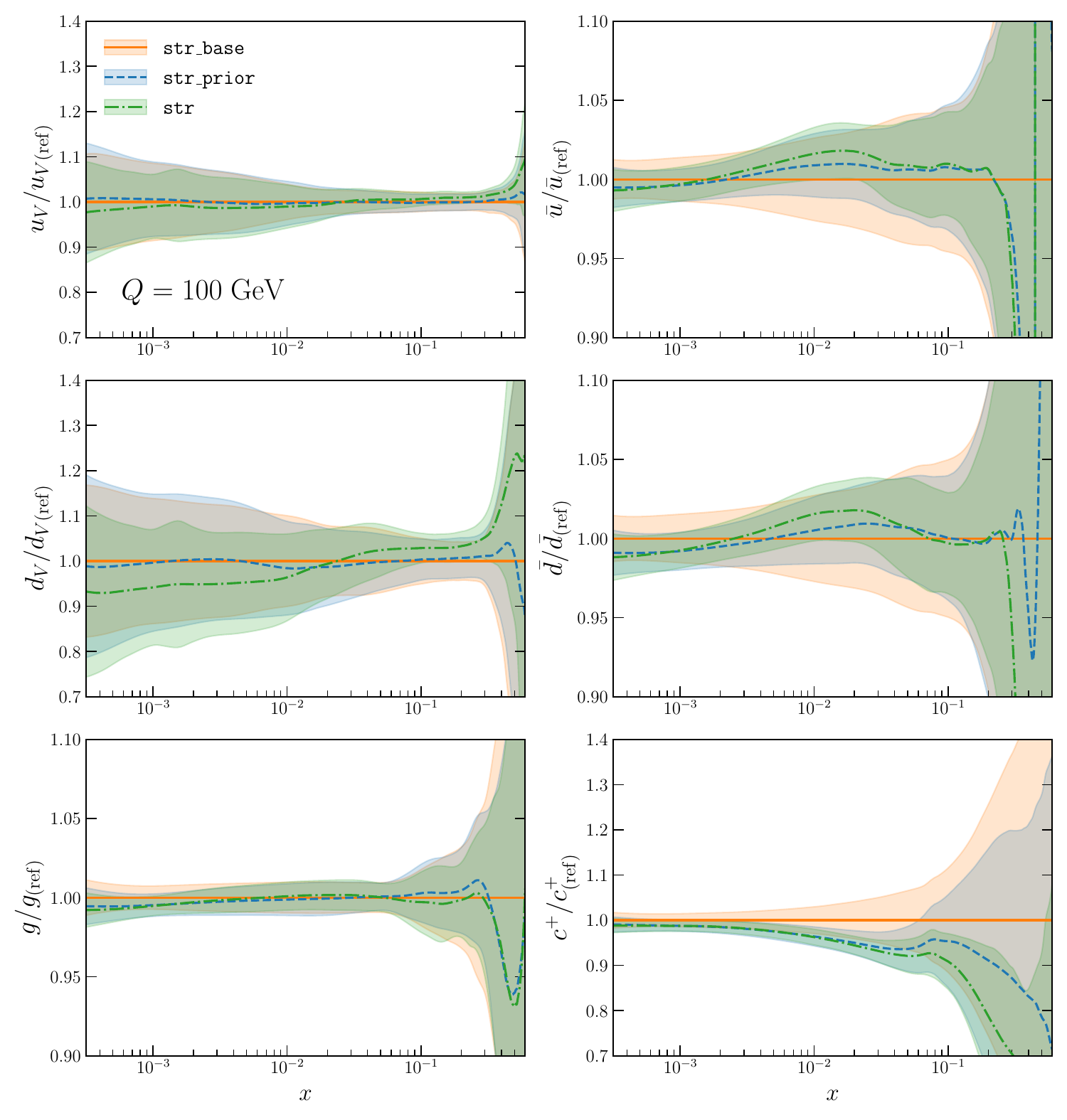}
 \caption{Comparison between the fits presented in this work.
   From top to bottom and left to right we show the up (valence and sea),
   down (valence and sea), gluon and charm
   distributions resulting from the {\tt str\_base}, {\tt str\_prior} and
   {\tt str} fits at $Q=100$~GeV. Results are normalized to the {\tt str\_base}
   fit (ref).}
\label{fig:pdfplot-nomad_allpdfs_q100gev}
\end{figure}

In Fig.~\ref{fig:pdfplot-nomad_allpdfs_q100gev} we compare the up (valence and
sea), down (valence and sea), gluon and charm distributions resulting from
the {\tt str\_base}, {\tt str\_prior} and {\tt str} fits at $Q=100$~GeV.
Results are normalized to the {\tt str\_base} fit.

From these comparisons, we observe that the new datasets have a little impact
on the gluon PDF, both on central values and on uncertainties, as expected. A
bigger effect is observed instead on the quark PDFs. For light quarks and
antiquarks, the electroweak LHC datasets constrain the distributions at
low to mid values of $x$, $x\lsim 0.1$, while the NOMAD datasets
do so at larger values of $x$, $x\gsim 0.1$. The two datasets are therefore
complementary, and concur together to enhance the central value of the down
distributions in the region $0.01\lsim x \sim 0.1$ by a few percent, and to
make all the light valence and sea quark PDFs more precise: overall,
uncertainties are reduced by up to a factor 2 in the same region for the
{\tt str} fit. For the charm PDF, the central value is suppressed in the
{\tt str} fit; uncertainties are reduced by up to a factor 2 for
$x\simeq 0.05$. This effect is almost entirely due to the NOMAD data, which
is indirectly sensitive to the charm PDF through its interplay with the
$s\bar{c}$ and  $\bar{s}c$ contributions to $W$-boson production.

\subsection{The strange content of the proton revisited}
\label{sec:strange}

We finally revisit the strange content of the proton in light of our results.
To this purpose, we consider the strange fraction of proton quark sea and the
corresponding ratio of momentum fraction, respectively defined as
\begin{equation}
  R_s(x,Q^2)
  =
  \frac{s(x,Q^2)+\bar{s}(x,Q^2)}{\bar{u}(x,Q^2)+\bar{d}(x,Q^2)}
  \,,
  \qquad
  K_s(Q^2)
  =
  \frac{\int_0^1 dx\, x \left[s(x,Q^2)+\bar{s}(x,Q^2)\right]}{\int_0^1 dx\, x \left[\bar{u}(x,Q^2)+\bar{d}(x,Q^2) \right]}\,.
  \label{eq:fractions}
\end{equation}

\begin{figure}[!t]
  \centering
\includegraphics[width=.49\textwidth]{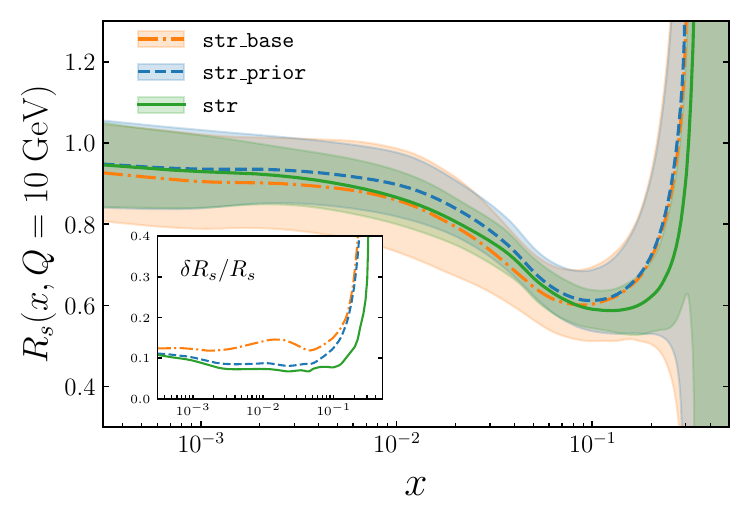}
\includegraphics[width=.49\textwidth]{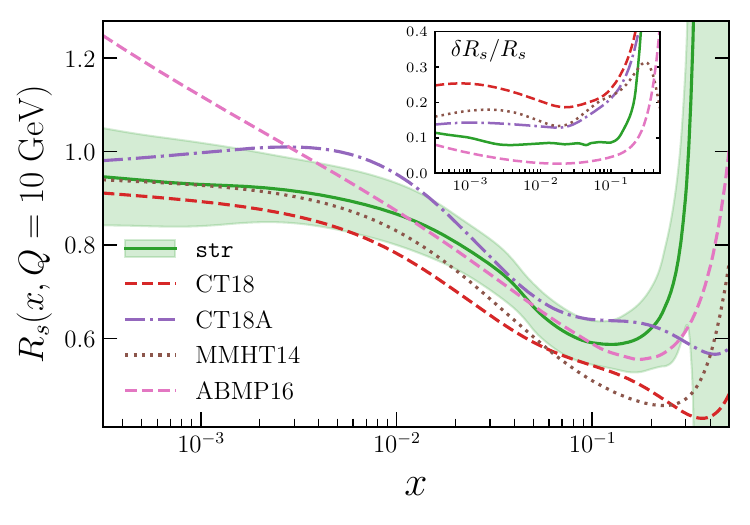}\\
\caption{The ratio $R_s$, Eq.~\eqref{eq:fractions}, as a function of $x$ at
  $Q=10$~GeV. The PDF used are from the {\tt str\_base}, {\tt str\_prior} and
  {\tt str} fits (left), and from the {\tt str} fit and from recent
  parton sets (right) see text for details. The insets display the
  corresponding relative uncertainty $\delta R_s/R_s$.}
\label{fig:Rs}
\end{figure}

We first consider the ratio $R_s$. In the left panel of Fig.~\ref{fig:Rs} we
display it for the {\tt str\_base}, {\tt str\_prior} and {\tt str} fits at a
scale $Q=10$~GeV as a function of $x$. The inset displays the associated
relative PDF uncertainty $\delta R_s/R_s$. The impact of the new datasets is
clearly visible. Concerning the central value, collider datasets do not alter
its expectation (the results obtained from the {\tt str\_base} and
{\tt str\_prior} fits are almost identical); the NOMAD dataset, instead,
prefers a more suppressed strange sea for $x\gsim 0.1$. Concerning
uncertainties, collider datasets lead to a reduction of the relative uncertainty
on $R_s$ of about 4\% for $x\lsim 0.1$; the NOMAD dataset, instead, reduces it
by about a factor of two for $x\gsim 0.1$. Overall, the impact of the new
datasets depends on $x$, and is mostly significant for $x= 0.2$, where the
uncertainty on $R_s$ is reduced from 20\% to 8\%. For $x\gsim 0.3$ no
experimental constraints are available, hence the PDF uncertainty blows up.

\begin{figure}[!t]
  \centering
  \includegraphics[width=.49\textwidth]{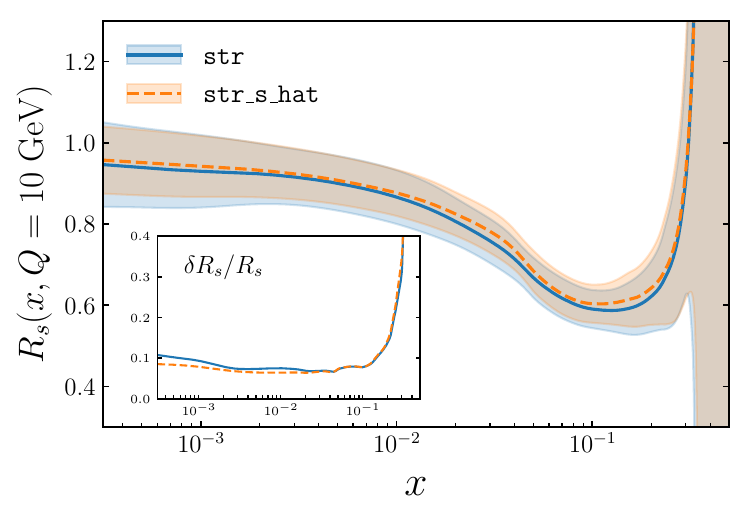}
  \includegraphics[width=.49\textwidth]{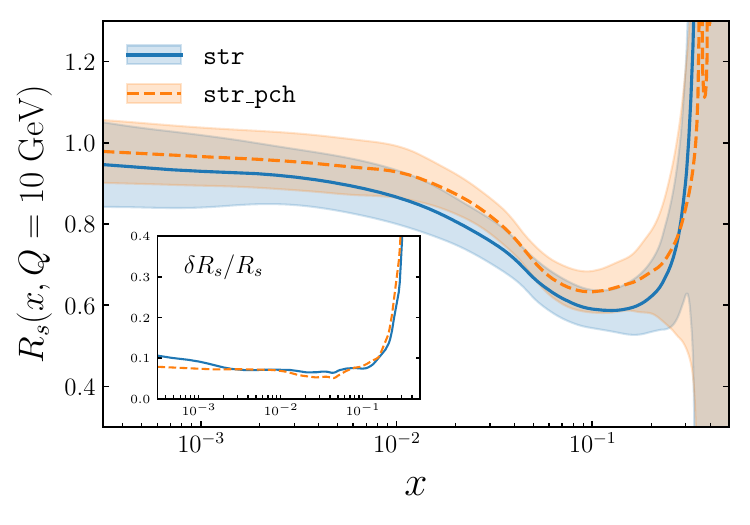}\\
  \caption{Same as Fig.~\ref{fig:Rs}, comparing fits to the
    $E_\nu$--dependent ({\tt str}) or to the $\sqrt{\hat{s}}$--dependent
    ({\tt str\_s\_hat}) NOMAD dataset (left), and fits with fitted
    ({\tt str}) or perturbative ({\tt str\_pch}) charm (right).}
  \label{fig:Rs_bis}
\end{figure}

The right panel of Fig.~\ref{fig:Rs} compares the ratio $R_s$, computed
at a scale $Q=10$~GeV as a function of $x$, as obtained from the {\tt str} fit
and from the CT18/CT18A~\cite{Hou:2019efy}, MMHT14~\cite{Harland-Lang:2014zoa},
and ABMP16~\cite{Alekhin:2017kpj} fits. The inset displays the
relative PDF uncertainty $\delta R_s/R_s$. Our {\tt str} determination agrees
with the CT18A and ABMP16 results within uncertainties in the data region.
However it overshoots the CT18 and MMHT14 results.
Note that the very small PDF uncertainties of
the ABMP16 result should be realistically rescaled by a tolerance factor
$T=\chi^2>1$~\cite{Gao:2017yyd}, which is however not accounted for in their
analysis. With this caveat, our results for $s^+$ and $R_s$ are also
the most precise, in particular around $x\sim 0.1$, thanks to the
wider dataset (and specifically of NOMAD) employed to constrain the
strange quark and anti-quark PDFs. 

We explicitly assessed how the results for $R_s$ obtained with our optimal fit {\tt str}
depend on the specific choice of the NOMAD dataset and on the fact that the
charm PDF is parametrized on the same footing as light quark PDFs.
In Fig.~\ref{fig:Rs_bis} we display the ratio $R_s$, as a function of $x$ at
$Q=10$~GeV: in the left panel we compare results obtained with the {\tt str}
and {\tt str\_s\_hat} fits; in the right panel, we compare results obtained
with the {\tt str} and {\tt str\_pch} fits. In the first case, both the
central value and the PDF uncertainties of $R_s$ are very similar. This fact
confirms the independence of our results upon the choice of the NOMAD dataset
included in the fit. In the second case, while PDF uncertainties turn out to
be very similar in both the perturbative and in the fitted charm fits, the
former prefers a central value which is systematically larger than the one
obtained from the latter. The size of the shift, however, is at most as large
as one-sigma in units of the PDF uncertainties, in line with previous
studies~\cite{Ball:2017nwa}.

\begin{figure}[!t]
\centering 
\includegraphics[width=.49\textwidth]{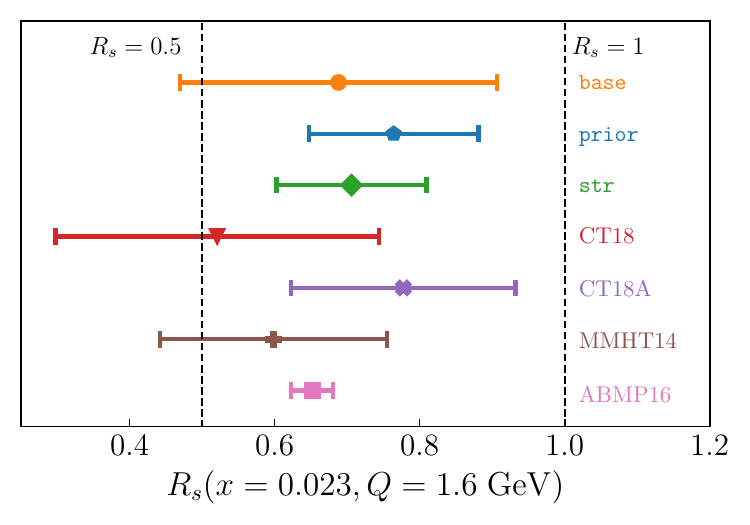}
\includegraphics[width=.49\textwidth]{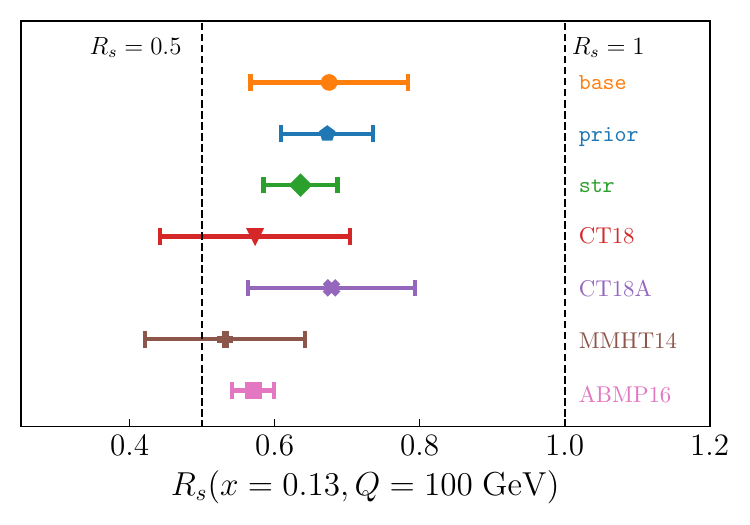}
\caption{The ratio $R_s$, Eq.~\eqref{eq:fractions}, computed at $x=0.023$,
  $Q=1.6$~GeV (left) and $x=0.13$, $Q=100$~GeV (right). The PDF sets
  are from the {\tt str\_base}, {\tt str\_prior}, {\tt str} fits and from
  other recent PDF analyses, see the text for details.}
\label{fig:rs-x-Q}
\end{figure}

In order to further investigate how our results compare to those reported
in the ATLAS studies~\cite{Aad:2012sb,Aaboud:2016btc}, which claimed a symmetric
strange quark sea, in Fig.~\ref{fig:rs-x-Q} we display the values of $R_s$ for
the {\tt base}, {\tt prior} and {\tt str} fits and for the fits shown in
Fig.~\ref{fig:Rs}. Here $R_s$ is evaluated for the two kinematic choices
outlined in~\cite{Aad:2012sb,Aaboud:2016btc}, first $x=0.023$ and $Q=1.6$~GeV,
and second for $x=0.13$ and $Q=100$~GeV. Fig.~\ref{fig:rs-x-Q} makes it clear
the consistent effect of the new datasets included in our analysis.
Considering the results for $R_s$ at $x=0.023$ and $Q=1.6$~GeV, the value
of  $R_s=0.69 \pm 0.22$
in the {\tt str\_base} fit is made more precise by the
LHC datasets, which reduce its uncertainty by about  a factor two,
while also increasing
  its central value, $R_s=0.76\pm 0.12$; then the neutrino-DIS NOMAD dataset
shifts this number towards a lower value by a half-sigma bringing in also
a further moderate reduction of the uncertainty, $R_s=0.71\pm 0.10$.
We therefore conclude that the result $R_s=1.13 \pm 0.11$, reported
in~\cite{Aaboud:2016btc} from an analysis of HERA and ATLAS $W$, $Z$ data
within the {\sc xFitter} framework~\cite{Alekhin:2014irh}, is not compatible
with ours, possibly because it is affected by a restricted
dataset and/or methodological limitations. 
Similar remarks also apply to the results for $R_s$ at $x=0.13$ and
$Q=100$~GeV, for which we observe a consistent slight reduction of the
central value of $R_s$, and a larger reduction of the uncertainty due to the
stronger effect of NOMAD data at larger $x$, see Fig.~\ref{fig:Rs}. Our
results are compatible, within uncertainties, with those of the other PDF determinations,
but are generally more precise.

\begin{figure}[!t]
\centering 
\includegraphics[width=.49\textwidth]{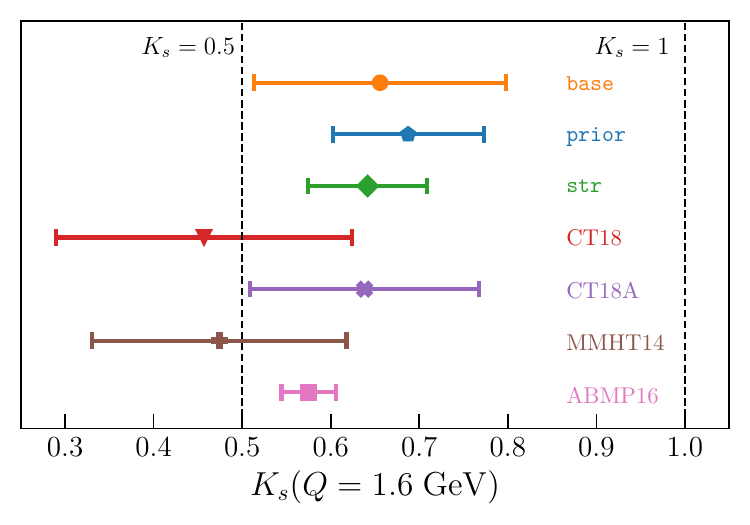}
\includegraphics[width=.49\textwidth]{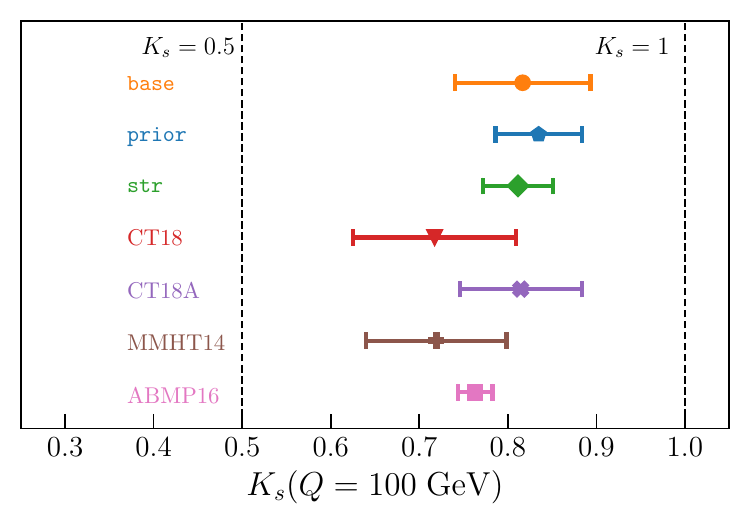}
\caption{The ratio $K_s$, Eq.~\eqref{eq:fractions}, for $Q=1.6$~GeV (left)
  and $Q=100$~GeV (right). The PDF sets are the same as in Fig.~\ref{fig:rs-x-Q}.}
\label{fig:ks}
\end{figure}

We finally consider the ratio of momentum fraction $K_s$, defined in
Eq.~\eqref{eq:fractions}.
Fig.~\ref{fig:ks} displays $K_s$ for $Q=1.6$~GeV and $Q=100$~GeV, respectively. The
qualitative interpretation of this quantity is consistent with that of $R_s$,
in particular, PDF uncertainties are reduced by a factor of two in the
{\tt str} fit with respect to the {\tt str\_base} fit. The values of $K_s$ grow
with the scale $Q$, as expected due to DGLAP evolution effects: for instance,
using the {\tt str} fit, one finds $K_s=0.64\pm 0.07$ at $Q=1.6$~GeV,
and $K_s=0.81\pm 0.04$ at $Q=100$~GeV. Overall, our final {\tt str} result
indicates that the strange sea is mildly suppressed with respect ot the rest
of the light sea quarks. The value of $K_s$ lies halfway a highly suppressed
($K_s\sim 0.5$) and a purely symmetric ($K_s=1$) scenarios. As for $R_s$,
our final {\tt str} result for $K_s$ is in agreement with the
determinations 
obtained by other recent PDF analyses within uncertainties, although
our results are generally more precise.

\section{Summary}
\label{sec:summary}

By means of a state-of-the-art global analysis, which combines all the relevant
experimental and theoretical inputs, we have achieved a precise determination
of the strangeness content of the proton. We have demonstrated the
compatibility of a wide range of strangeness-sensitive datasets; quantified
their relative impact on the fit; compared our results to other recent global
analyses; and assessed the robustness of our results with respect to various
methodological choices. Our analysis demonstrates that the strange PDF can be
precisely determined and that, after all, the proton is not too {\it strange}:
the momentum fraction carried by strange quark and antiquark PDFs ranges between
about 65\% and 80\% of the momentum fraction carried by the other light sea
quarks in a wide energy range (1.6~GeV $\leq Q\leq 100$~GeV).
{  The present determination of the strangeness content
  of the proton is found to agree, within uncertainties, with
the results of other recent global PDF analyses.}

Pivotal to this result is the complementary between the LHC gauge boson production
data and of the charmed-tagged neutrino-DIS data, in particular from the NOMAD
experiment. Our {\tt str} PDF set, which combines all this information, is
available in the {\sc LHAPDF} format~\cite{Buckley:2014ana} together with its
perturbative charm counterpart from
\begin{center}
  \tt \href{http://nnpdf.mi.infn.it/nnpdf3-1strangeness/}{http://nnpdf.mi.infn.it/nnpdf3-1strangeness/}
  \end{center}
This analysis represents an important input for phenomenology, for instance to carry out
improved determinations of fundamental parameters of the SM or to be used as
baseline in the determination of nuclear PDFs, where strange distributions are
not well known~\cite{AbdulKhalek:2020yuc,Eskola:2016oht,Kusina:2020lyz}.
Our determination of the strange and anti-strange quark PDFs
could  be further stress-tested with more exclusive processes,
{\it e.g.}, measurements of kaon production in semi-inclusive DIS (SIDIS).
Studies of the strange PDFs based on SIDIS~\cite{Airapetian:2013zaw,
  Borsa:2017vwy,Sato:2019yez} notoriously prefer a suppressed strangeness,
but are also subject to the potential bias coming from their sensitivity to the
fragmentation of the strange quarks into kaons.

\subsection*{Acknowledgments}

We are grateful to Jun Gao for providing us with the NNLO $K$-factors
for the NuTeV and the NOMAD measurements, and for help into the benchmark of
the corresponding NLO calculations. We thank Valerio Bertone for assistance
with the usage of {\sc APFEL}, Lucian Harland-Lang for discussions on the
massive corrections to dimuon production,
and Rabah Abdul-Khalek, Amanda Cooper-Sarkar, Stefano Forte, Katerina Lipka,
Gavin Pownall and Cameron Voisey for comments on the manuscript.
E.R.N.\ is supported by the European Commission through the Marie
Sk\l odowska-Curie Action ParDHonS FFs.TMDs (grant number 752748).
J.R.\ is partially supported by the Netherlands Organization for Scientific
Research (NWO).
M.U.\ and S.I.\ are partially supported by the STFC grant ST/L000385/1 and by
the Royal Society grant RGF/EA/180148.
The work of M.U.\ is also funded by the Royal Society grant DH150088
and supported by the European Research Council under the 
European Union’s Horizon 2020 research and innovation Programme (grant agreement n.950246). 

\providecommand{\href}[2]{#2}\begingroup\raggedright\endgroup


\begin{thebibliography}{10}

\bibitem{Gao:2017yyd}
J.~Gao, L.~Harland-Lang, and J.~Rojo, {\it {The Structure of the Proton in the
  LHC Precision Era}},  {\em Phys. Rept.} {\bf 742} (2018) 1--121,
  [\href{http://arxiv.org/abs/1709.04922}{{\tt arXiv:1709.04922}}].

\bibitem{Kovarik:2019xvh}
K.~Kova\v{r}\'\i{}k, P.~M. Nadolsky, and D.~E. Soper, {\it {Hadron structure in
  high-energy collisions}},  {\em Rev. Mod. Phys.} {\bf 92} (2020), no.~4
  045003, [\href{http://arxiv.org/abs/1905.06957}{{\tt arXiv:1905.06957}}].

\bibitem{Ethier:2020way}
J.~J. Ethier and E.~R. Nocera, {\it {Parton Distributions in Nucleons and
  Nuclei}},  {\em Ann.\ Rev.\ Nucl.\ Part.\ Sci.} (2020), no.~70 1--34,
  [\href{http://arxiv.org/abs/2001.07722}{{\tt arXiv:2001.07722}}].

\bibitem{Aaboud:2017svj}
{\bf ATLAS} Collaboration, M.~Aaboud et~al., {\it {Measurement of the $W$-boson
  mass in pp collisions at $\sqrt{s}=7$ TeV with the ATLAS detector}},  {\em
  Eur. Phys. J.} {\bf C78} (2018), no.~2 110,
  [\href{http://arxiv.org/abs/1701.07240}{{\tt arXiv:1701.07240}}].

\bibitem{Sirunyan:2018swq}
{\bf CMS} Collaboration, A.~M. Sirunyan et~al., {\it {Measurement of the weak
  mixing angle using the forward-backward asymmetry of Drell-Yan events in pp
  collisions at 8 TeV}},  {\em Eur. Phys. J.} {\bf C78} (2018), no.~9 701,
  [\href{http://arxiv.org/abs/1806.00863}{{\tt arXiv:1806.00863}}].

\bibitem{Bagnaschi:2019mzi}
E.~Bagnaschi and A.~Vicini, {\it {A new look at the estimation of the PDF
  uncertainties in the determination of electroweak parameters at hadron
  colliders}},  \href{http://arxiv.org/abs/1910.04726}{{\tt arXiv:1910.04726}}.

\bibitem{Bazarko:1994tt}
{\bf CCFR} Collaboration, A.~O. Bazarko et~al., {\it {Determination of the
  strange quark content of the nucleon from a next-to-leading order QCD
  analysis of neutrino charm production}},  {\em Z. Phys.} {\bf C65} (1995)
  189--198, [\href{http://arxiv.org/abs/hep-ex/9406007}{{\tt hep-ex/9406007}}].

\bibitem{KayisTopaksu:2008aa}
{\bf CHORUS} Collaboration, A.~Kayis-Topaksu et~al., {\it {Leading order
  analysis of neutrino induced dimuon events in the CHORUS experiment}},  {\em
  Nucl. Phys. B} {\bf 798} (2008) 1--16,
  [\href{http://arxiv.org/abs/0804.1869}{{\tt arXiv:0804.1869}}].

\bibitem{Mason:2007zz}
D.~Mason et~al., {\it {Measurement of the Nucleon Strange-Antistrange Asymmetry
  at Next-to-Leading Order in QCD from NuTeV Dimuon Data}},  {\em Phys. Rev.
  Lett.} {\bf 99} (2007) 192001.

\bibitem{Samoylov:2013xoa}
{\bf NOMAD} Collaboration, O.~Samoylov et~al., {\it {A Precision Measurement of
  Charm Dimuon Production in Neutrino Interactions from the NOMAD Experiment}},
   {\em Nucl.Phys.} {\bf B876} (2013) 339,
  [\href{http://arxiv.org/abs/1308.4750}{{\tt arXiv:1308.4750}}].

\bibitem{Aad:2012sb}
{\bf ATLAS} Collaboration, G.~Aad et~al., {\it {Determination of the strange
  quark density of the proton from ATLAS measurements of the $W,Z$ cross
  sections}},  {\em Phys.Rev.Lett.} (2012)
  [\href{http://arxiv.org/abs/1203.4051}{{\tt arXiv:1203.4051}}].

\bibitem{Aaboud:2016btc}
{\bf ATLAS} Collaboration, M.~Aaboud et~al., {\it {Precision measurement and
  interpretation of inclusive $W^+$ , $W^-$ and $Z/\gamma ^*$ production cross
  sections with the ATLAS detector}},  {\em Eur. Phys. J.} {\bf C77} (2017),
  no.~6 367, [\href{http://arxiv.org/abs/1612.03016}{{\tt arXiv:1612.03016}}].

\bibitem{Sutton:2019pug}
{\bf ATLAS} Collaboration, M.~Sutton, {\it {The PDF interpretation of the
  measurement of a vector boson produced in association with jets at the ATLAS
  detector}},  {\em PoS} {\bf DIS2019} (2019) 034.

\bibitem{Stirling:2012vh}
W.~Stirling and E.~Vryonidou, {\it {Charm production in association with an
  electroweak gauge boson at the LHC}},  {\em Phys.Rev.Lett.} {\bf 109} (2012)
  082002, [\href{http://arxiv.org/abs/1203.6781}{{\tt arXiv:1203.6781}}].

\bibitem{Aaboud:2017soa}
{\bf ATLAS} Collaboration, M.~Aaboud et~al., {\it {Measurement of differential
  cross sections and $W^+/W^-$ cross-section ratios for $W$ boson production in
  association with jets at $\sqrt{s}=8$ TeV with the ATLAS detector}},  {\em
  JHEP} {\bf 05} (2018) 077, [\href{http://arxiv.org/abs/1711.03296}{{\tt
  arXiv:1711.03296}}].

\bibitem{Aad:2014xca}
{\bf ATLAS} Collaboration, G.~Aad et~al., {\it {Measurement of the production
  of a $W$ boson in association with a charm quark in $pp$ collisions at
  $\sqrt{s} =$ 7 TeV with the ATLAS detector}},  {\em JHEP} {\bf 1405} (2014)
  068, [\href{http://arxiv.org/abs/1402.6263}{{\tt arXiv:1402.6263}}].

\bibitem{Chatrchyan:2013uja}
{\bf CMS} Collaboration, S.~Chatrchyan et~al., {\it {Measurement of associated
  W + charm production in pp collisions at $\sqrt{s}$ = 7 TeV}},  {\em JHEP}
  {\bf 02} (2014) 013, [\href{http://arxiv.org/abs/1310.1138}{{\tt
  arXiv:1310.1138}}].

\bibitem{Sirunyan:2018hde}
{\bf CMS} Collaboration, A.~M. Sirunyan et~al., {\it {Measurement of associated
  production of a W boson and a charm quark in proton-proton collisions at
  $\sqrt{s} =$ 13 TeV}},  {\em Eur. Phys. J.} {\bf C79} (2019), no.~3 269,
  [\href{http://arxiv.org/abs/1811.10021}{{\tt arXiv:1811.10021}}].

\bibitem{Ball:2017nwa}
{\bf NNPDF} Collaboration, R.~D. Ball et~al., {\it {Parton distributions from
  high-precision collider data}},  {\em Eur. Phys. J.} {\bf C77} (2017), no.~10
  663, [\href{http://arxiv.org/abs/1706.00428}{{\tt arXiv:1706.00428}}].

\bibitem{Hou:2019efy}
T.-J. Hou et~al., {\it {New CTEQ global analysis of quantum chromodynamics with
  high-precision data from the LHC}},
  \href{http://arxiv.org/abs/1912.10053}{{\tt arXiv:1912.10053}}.

\bibitem{Thorne:2019mpt}
R.~S. Thorne, S.~Bailey, T.~Cridge, L.~A. Harland-Lang, A.~Martin, and
  R.~Nathvani, {\it {Updates of PDFs using the MMHT framework}},  {\em PoS}
  {\bf DIS2019} (2019) 036, [\href{http://arxiv.org/abs/1907.08147}{{\tt
  arXiv:1907.08147}}].

\bibitem{Lai:2007dq}
H.~L. Lai et~al., {\it {The Strange Parton Distribution of the Nucleon: Global
  Analysis and Applications}},  {\em JHEP} {\bf 04} (2007) 089,
  [\href{http://arxiv.org/abs/hep-ph/0702268}{{\tt hep-ph/0702268}}].

\bibitem{Kusina:2012vh}
A.~Kusina, T.~Stavreva, S.~Berge, F.~Olness, I.~Schienbein, K.~Kovarik,
  T.~Jezo, J.~Yu, and K.~Park, {\it {Strange Quark PDFs and Implications for
  Drell-Yan Boson Production at the LHC}},  {\em Phys. Rev. D} {\bf 85} (2012)
  094028, [\href{http://arxiv.org/abs/1203.1290}{{\tt arXiv:1203.1290}}].

\bibitem{Alekhin:2014sya}
S.~Alekhin, J.~Blumlein, L.~Caminadac, K.~Lipka, K.~Lohwasser, S.~Moch,
  R.~Petti, and R.~Placakyte, {\it {Determination of Strange Sea Quark
  Distributions from Fixed-target and Collider Data}},  {\em Phys. Rev.} {\bf
  D91} (2015), no.~9 094002, [\href{http://arxiv.org/abs/1404.6469}{{\tt
  arXiv:1404.6469}}].

\bibitem{Alekhin:2017olj}
S.~Alekhin, J.~Blümlein, and S.~Moch, {\it {Strange sea determination from
  collider data}},  {\em Phys. Lett. B} {\bf 777} (2018) 134--140,
  [\href{http://arxiv.org/abs/1708.01067}{{\tt arXiv:1708.01067}}].

\bibitem{Cooper-Sarkar:2018ufj}
A.~Cooper-Sarkar and K.~Wichmann, {\it {QCD analysis of the ATLAS and CMS
  $W^{\pm}$ and $Z$ cross-section measurements and implications for the strange
  sea density}},  {\em Phys. Rev. D} {\bf 98} (2018), no.~1 014027,
  [\href{http://arxiv.org/abs/1803.00968}{{\tt arXiv:1803.00968}}].

\bibitem{Ball:2018iqk}
{\bf NNPDF} Collaboration, R.~D. Ball, S.~Carrazza, L.~Del~Debbio, S.~Forte,
  Z.~Kassabov, J.~Rojo, E.~Slade, and M.~Ubiali, {\it {Precision determination
  of the strong coupling constant within a global PDF analysis}},  {\em Eur.
  Phys. J.} {\bf C78} (2018), no.~5 408,
  [\href{http://arxiv.org/abs/1802.03398}{{\tt arXiv:1802.03398}}].

\bibitem{Aaltonen:2010zza}
{\bf CDF} Collaboration, T.~A. Aaltonen et~al., {\it {Measurement of
  $d\sigma/dy$ of Drell-Yan $e^+e^-$ pairs in the $Z$ Mass Region from
  $p\bar{p}$ Collisions at $\sqrt{s}=1.96$ TeV}},  {\em Phys. Lett.} {\bf B692}
  (2010) 232--239, [\href{http://arxiv.org/abs/0908.3914}{{\tt
  arXiv:0908.3914}}].

\bibitem{Abazov:2007jy}
{\bf D0} Collaboration, V.~M. Abazov et~al., {\it {Measurement of the shape of
  the boson rapidity distribution for $p \bar{p} \to Z/\gamma^* \to e^{+}
  e^{-}$ + $X$ events produced at $\sqrt{s}$=1.96-TeV}},  {\em Phys. Rev.} {\bf
  D76} (2007) 012003, [\href{http://arxiv.org/abs/hep-ex/0702025}{{\tt
  hep-ex/0702025}}].

\bibitem{Aad:2011dm}
{\bf ATLAS} Collaboration, G.~Aad et~al., {\it {Measurement of the inclusive
  $W^{\pm}$ and $Z/\gamma^*$ cross sections in the electron and muon decay
  channels in pp collisions at $\sqrt{s}$= 7 TeV with the ATLAS detector}},
  {\em Phys.Rev.} {\bf D85} (2012) 072004,
  [\href{http://arxiv.org/abs/1109.5141}{{\tt arXiv:1109.5141}}].

\bibitem{Khachatryan:2016pev}
{\bf CMS} Collaboration, V.~Khachatryan et~al., {\it {Measurement of the
  differential cross section and charge asymmetry for inclusive $\mathrm
  {p}\mathrm {p}\rightarrow \mathrm {W}^{\pm }+X$ production at ${\sqrt{s}} =
  8$ TeV}},  {\em Eur. Phys. J.} {\bf C76} (2016), no.~8 469,
  [\href{http://arxiv.org/abs/1603.01803}{{\tt arXiv:1603.01803}}].

\bibitem{Bertone:2013vaa}
V.~Bertone, S.~Carrazza, and J.~Rojo, {\it {APFEL: A PDF Evolution Library with
  QED corrections}},  {\em Comput.Phys.Commun.} {\bf 185} (2014) 1647,
  [\href{http://arxiv.org/abs/1310.1394}{{\tt arXiv:1310.1394}}].

\bibitem{Gao:2017kkx}
J.~Gao, {\it {Massive charged-current coefficient functions in deep-inelastic
  scattering at NNLO and impact on strange-quark distributions}},  {\em JHEP}
  {\bf 02} (2018) 026, [\href{http://arxiv.org/abs/1710.04258}{{\tt
  arXiv:1710.04258}}].

\bibitem{Berger:2016inr}
E.~L. Berger, J.~Gao, C.~S. Li, Z.~L. Liu, and H.~X. Zhu, {\it {Charm-Quark
  Production in Deep-Inelastic Neutrino Scattering at Next-to-Next-to-Leading
  Order in QCD}},  {\em Phys.\ Rev.\ Lett.} {\bf 116} (2016), no.~21 212002,
  [\href{http://arxiv.org/abs/1601.05430}{{\tt arXiv:1601.05430}}].

\bibitem{Forte:2010ta}
S.~Forte, E.~Laenen, P.~Nason, and J.~Rojo, {\it {Heavy quarks in
  deep-inelastic scattering}},  {\em Nucl. Phys.} {\bf B834} (2010) 116--162,
  [\href{http://arxiv.org/abs/1001.2312}{{\tt arXiv:1001.2312}}].

\bibitem{Ball:2011mu}
{\bf {The NNPDF }} Collaboration, R.~D. Ball et~al., {\it {Impact of Heavy
  Quark Masses on Parton Distributions and LHC Phenomenology}},  {\em Nucl.
  Phys.} {\bf B849} (2011) 296, [\href{http://arxiv.org/abs/1101.1300}{{\tt
  arXiv:1101.1300}}].

\bibitem{Boughezal:2016wmq}
R.~Boughezal, J.~M. Campbell, R.~K. Ellis, C.~Focke, W.~Giele, X.~Liu,
  F.~Petriello, and C.~Williams, {\it {Color singlet production at NNLO in
  MCFM}},  {\em Eur. Phys. J. C} {\bf 77} (2017), no.~1 7,
  [\href{http://arxiv.org/abs/1605.08011}{{\tt arXiv:1605.08011}}].

\bibitem{Carli:2010rw}
T.~Carli et~al., {\it {A posteriori inclusion of parton density functions in
  NLO QCD final-state calculations at hadron colliders: The APPLGRID Project}},
   {\em Eur.Phys.J.} {\bf C66} (2010) 503,
  [\href{http://arxiv.org/abs/0911.2985}{{\tt arXiv:0911.2985}}].

\bibitem{Gavin:2010az}
R.~Gavin, Y.~Li, F.~Petriello, and S.~Quackenbush, {\it {FEWZ 2.0: A code for
  hadronic Z production at next-to-next-to-leading order}},  {\em Comput. Phys.
  Commun.} {\bf 182} (2011) 2388--2403,
  [\href{http://arxiv.org/abs/1011.3540}{{\tt arXiv:1011.3540}}].

\bibitem{Boughezal:2015dva}
R.~Boughezal, C.~Focke, X.~Liu, and F.~Petriello, {\it {$W$-boson production in
  association with a jet at next-to-next-to-leading order in perturbative
  QCD}},  {\em Phys. Rev. Lett.} {\bf 115} (2015), no.~6 062002,
  [\href{http://arxiv.org/abs/1504.02131}{{\tt arXiv:1504.02131}}].

\bibitem{Ridder:2015dxa}
A.~Gehrmann-De~Ridder, T.~Gehrmann, E.~W.~N. Glover, A.~Huss, and T.~A. Morgan,
  {\it {Precise QCD predictions for the production of a Z boson in association
  with a hadronic jet}},  {\em Phys. Rev. Lett.} {\bf 117} (2016), no.~2
  022001, [\href{http://arxiv.org/abs/1507.02850}{{\tt arXiv:1507.02850}}].

\bibitem{Czakon:2020coa}
M.~Czakon, A.~Mitov, M.~Pellen, and R.~Poncelet, {\it {NNLO QCD predictions for
  W+c-jet production at the LHC}},  \href{http://arxiv.org/abs/2011.01011}{{\tt
  arXiv:2011.01011}}.

\bibitem{AbdulKhalek:2019ihb}
{\bf NNPDF} Collaboration, R.~Abdul~Khalek et~al., {\it {Parton Distributions
  with Theory Uncertainties: General Formalism and First Phenomenological
  Studies}},  {\em Eur.\ Phys.\ J.\ C} {\bf 79} (2019), no.~11 931,
  [\href{http://arxiv.org/abs/1906.10698}{{\tt arXiv:1906.10698}}].

\bibitem{AbdulKhalek:2019bux}
{\bf NNPDF} Collaboration, R.~Abdul~Khalek et~al., {\it {A first determination
  of parton distributions with theoretical uncertainties}},  {\em Eur.\ Phys.\
  J.} {\bf C} (2019) 79:838, [\href{http://arxiv.org/abs/1905.04311}{{\tt
  arXiv:1905.04311}}].

\bibitem{Ball:2018twp}
{\bf NNPDF} Collaboration, R.~D. Ball, E.~R. Nocera, and R.~L. Pearson, {\it
  {Nuclear Uncertainties in the Determination of Proton PDFs}},  {\em Eur.
  Phys. J.} {\bf C79} (2019), no.~3 282,
  [\href{http://arxiv.org/abs/1812.09074}{{\tt arXiv:1812.09074}}].

\bibitem{AbdulKhalek:2020yuc}
R.~Abdul~Khalek, J.~J. Ethier, J.~Rojo, and G.~van Weelden, {\it {nNNPDF2.0:
  quark flavor separation in nuclei from LHC data}},  {\em JHEP} {\bf 09}
  (2020) 183, [\href{http://arxiv.org/abs/2006.14629}{{\tt arXiv:2006.14629}}].

\bibitem{Ball:2014uwa}
{\bf NNPDF} Collaboration, R.~D. Ball et~al., {\it {Parton distributions for
  the LHC Run II}},  {\em JHEP} {\bf 04} (2015) 040,
  [\href{http://arxiv.org/abs/1410.8849}{{\tt arXiv:1410.8849}}].

\bibitem{Ball:2011gg}
R.~D. Ball, V.~Bertone, F.~Cerutti, L.~Del~Debbio, S.~Forte, et~al., {\it
  {Reweighting and Unweighting of Parton Distributions and the LHC W lepton
  asymmetry data}},  {\em Nucl.Phys.} {\bf B855} (2012) 608--638,
  [\href{http://arxiv.org/abs/1108.1758}{{\tt arXiv:1108.1758}}].

\bibitem{Ball:2010gb}
{\bf The NNPDF} Collaboration, R.~D. Ball et~al., {\it {Reweighting NNPDFs: the
  W lepton asymmetry}},  {\em Nucl. Phys.} {\bf B849} (2011) 112--143,
  [\href{http://arxiv.org/abs/1012.0836}{{\tt arXiv:1012.0836}}].

\bibitem{Nocera:2019wyk}
E.~R. Nocera, M.~Ubiali, and C.~Voisey, {\it {Single Top Production in PDF
  fits}},  {\em JHEP} {\bf 05} (2020) 067,
  [\href{http://arxiv.org/abs/1912.09543}{{\tt arXiv:1912.09543}}].

\bibitem{Harland-Lang:2014zoa}
L.~A. Harland-Lang, A.~D. Martin, P.~Motylinski, and R.~S. Thorne, {\it {Parton
  distributions in the LHC era: MMHT 2014 PDFs}},  {\em Eur. Phys. J.} {\bf
  C75} (2015) 204, [\href{http://arxiv.org/abs/1412.3989}{{\tt
  arXiv:1412.3989}}].

\bibitem{Alekhin:2017kpj}
S.~Alekhin, J.~Blümlein, S.~Moch, and R.~Placakyte, {\it {Parton distribution
  functions, $\alpha_s$, and heavy-quark masses for LHC Run II}},  {\em Phys.
  Rev.} {\bf D96} (2017), no.~1 014011,
  [\href{http://arxiv.org/abs/1701.05838}{{\tt arXiv:1701.05838}}].

\bibitem{Ball:2016neh}
{\bf NNPDF} Collaboration, R.~D. Ball, V.~Bertone, M.~Bonvini, S.~Carrazza,
  S.~Forte, A.~Guffanti, N.~P. Hartland, J.~Rojo, and L.~Rottoli, {\it {A
  Determination of the Charm Content of the Proton}},  {\em Eur. Phys. J.} {\bf
  C76} (2016), no.~11 647, [\href{http://arxiv.org/abs/1605.06515}{{\tt
  arXiv:1605.06515}}].

\bibitem{Alekhin:2014irh}
S.~Alekhin et~al., {\it {HERAFitter}},  {\em Eur. Phys. J.} {\bf C75} (2015),
  no.~7 304, [\href{http://arxiv.org/abs/1410.4412}{{\tt arXiv:1410.4412}}].

\bibitem{Buckley:2014ana}
A.~Buckley, J.~Ferrando, S.~Lloyd, K.~Nordström, B.~Page, et~al., {\it
  {LHAPDF6: parton density access in the LHC precision era}},  {\em
  Eur.Phys.J.} {\bf C75} (2015) 132,
  [\href{http://arxiv.org/abs/1412.7420}{{\tt arXiv:1412.7420}}].

\bibitem{Eskola:2016oht}
K.~J. Eskola, P.~Paakkinen, H.~Paukkunen, and C.~A. Salgado, {\it {EPPS16:
  Nuclear parton distributions with LHC data}},  {\em Eur. Phys. J. C} {\bf 77}
  (2017), no.~3 163, [\href{http://arxiv.org/abs/1612.05741}{{\tt
  arXiv:1612.05741}}].

\bibitem{Kusina:2020lyz}
A.~Kusina et~al., {\it {Impact of LHC vector boson production in heavy ion
  collisions on strange PDFs}},  \href{http://arxiv.org/abs/2007.09100}{{\tt
  arXiv:2007.09100}}.

\bibitem{Airapetian:2013zaw}
{\bf HERMES} Collaboration, A.~Airapetian et~al., {\it {Reevaluation of the
  parton distribution of strange quarks in the nucleon}},  {\em Phys. Rev.}
  {\bf D89} (2014), no.~9 097101, [\href{http://arxiv.org/abs/1312.7028}{{\tt
  arXiv:1312.7028}}].

\bibitem{Borsa:2017vwy}
I.~Borsa, R.~Sassot, and M.~Stratmann, {\it {Probing the Sea Quark Content of
  the Proton with One-Particle-Inclusive Processes}},  {\em Phys. Rev.} {\bf
  D96} (2017), no.~9 094020, [\href{http://arxiv.org/abs/1708.01630}{{\tt
  arXiv:1708.01630}}].

\bibitem{Sato:2019yez}
{\bf JAM} Collaboration, N.~Sato, C.~Andres, J.~Ethier, and W.~Melnitchouk,
  {\it {Strange quark suppression from a simultaneous Monte Carlo analysis of
  parton distributions and fragmentation functions}},  {\em Phys. Rev. D} {\bf
  101} (2020), no.~7 074020, [\href{http://arxiv.org/abs/1905.03788}{{\tt
  arXiv:1905.03788}}].

\end{thebibliography}
\end{document}